\def\ispreprint{1}  % turn to zero if you want for submission to Icarus, turn to 1 if you want a preprint style
 
\if \ispreprint1
\documentclass[preprint,authoryear,5p]{elsarticle}  % preprint style
\else
\documentclass[preprint,authoryear,12pt]{elsarticle}  %  for submission
\fi

\usepackage[utf8]{inputenc}
\usepackage{lineno} % needed for submission to Icarus
\usepackage{amssymb}
\usepackage{amsmath,bm}
\usepackage{graphicx} % Required for inserting images
\usepackage[draft]{hyperref} % better urls?
\usepackage{multicol}
\usepackage{natbib} % biblio
\usepackage{threeparttable}

\begin{document}

\title{The azimuthal distribution of ejecta mass from oblique impacts into sand}

\author[a1]{Alice C. Quillen\corref{cor1}}
\ead{alice.quillen@rochester.edu}

\author[a1]{Sean Doran}
\ead{sdoran3@u.rochester.edu}

%\author[a1]{Luke O'Brient?}
%\ead{lobrient@u.rochester.edu}

%\author[a2]{Mokin Lee?}
%\ead{mlee80@u.rochester.edu}

%\author{Stephen Luniewski?}
%\ead{sluniews@ur.rochester.edu}

%\author{Maggie Ju?}
%\ead{mju2@u.rochester.edu}

%\author{List and order TBA}

\address[a1]{Department of Physics and Astronomy, University of Rochester, Rochester, NY 14627, USA}

%\address[a2]{Department of Mechanical Engineering, University of Rochester, Rochester, NY 14627, USA}
%\address[a2]{Department of Earth and Environmental Science, University of Rochester, Rochester, NY 14627, USA}

%\date{January 2024}

\cortext[cor1]{Corresponding author}

\begin{abstract}
We measure ejecta mass as a function of azimuthal and impact angle for 104 m/s oblique impacts into sand.  We find that the ejecta mass distribution is strongly sensitive to azimuthal angle with as high as 8 times more mass in ejecta on the downrange side compared to the uprange side. Crater radii,  measured from the impact point, are measured at different impact and azimuthal angles.  Crater ejecta scaling laws are modified to depend on azimuthal and impact angle. We find that crater radii are sensitive to both impact and azimuthal angle but the ejecta mass as a function of both angles can be estimated from the cube of the crater radius without an additional angular dependent function.  The ejecta distributions are relevant for processes that depend upon the integrated properties of approximately 100 m/s impacts occurring in the outer solar system and possibly during planetesimal formation. 

\end{abstract}

% target journal  Meteoritics and Planetary Science (Wiley)
% https://onlinelibrary.wiley.com/page/journal/19455100/homepage/forauthors.html

%\begin{document}

\maketitle
\if\ispreprint1
\else
\linenumbers %  turn this on for submission to Icarus
\fi

\section{Introduction}

Because ejecta from an impact can escape an asteroid or planetesimal, the velocity and mass distributions of ejecta affect the total momentum imparted to the asteroid and as a consequence, its deflected trajectory \citep{Jutzi_2014,Cheng_2016,Thomas_2023}.    
The ratio of the change in center of mass momentum and the projectile momentum is known as 
the momentum transfer efficiency parameter, and is relevant for mitigation strategies for deflection of potentially hazardous asteroids \citep{Holsapple_2012}.  
Over a long period of time, impacts cumulatively exert a torque on a spinning body, which can reduce its rotation rate \citep{Dobrovolskis_1984,Luniewski_2024}.  
The distribution of ejecta in a lower velocity impact regime (10 to 50 m/s) is relevant for estimating 
collision evolution of asteroids \citep{Farinella_1992,Campo_1994}, accretion, erosion and particle transport on young and forming planetesimals \citep{Quillen_2024}. 

%Thus, ejection dynamics have consequences for early planetary accretion, and collisional evolution of asteroids (e.g., Farinella and Davis 1992; Campo Bagatin et al. 1994).
%The total momentum in impact generated ejecta affects estimates for asteroid spin evolution. 

%The Double Asteroid Redirection Test (DART) mission \citep{Rivkin_2021} successfully measured the momentum transmitted to an asteroid following a 6.6 km/s impact \citep{Thomas_2023}. 
%The DART momentum transfer efficiency parameter places constraints on the constituents and structure of the target  asteroid \citep{Thomas_2023}.  

While most craters on asteroid and satellite surfaces, including the Moon, are nearly round, impacts on astronomical bodies rarely have projectile velocity vector nearly normal to the surface. 
We take the impact angle $\theta_I$ to be the angle between the plane tangent to the surface at the point of impact and the projectile momentum vector, as seen in a reference frame in which the surface is stationary.  With this convention, grazing impacts have low impact angle. 
More than 50\% of astronomical impacts are expected to occur at impact angles between 30 and $60^\circ$ \citep{Pierazzo_2000}.   High velocity (up to 7.2 km/s) laboratory impacts into quarz sand, pumice powder and granite substrates, give round craters for impact angles above about $20^\circ$ (\citealt{Gault_1978}; see their Figures 1 and 2).   
%Unless the impact angle is below $20 to 30^\circ$, crater shape is not sensitive to impact angle.  
Laboratory experiments show that 
ejecta deposits and curtains can be asymmetric at a higher impact angle than $30^\circ$ 
\citep{Gault_1978,Schultz_1999,Anderson_2003,Hessen_2007,Suo_2024}.
For example, 
Figure 18 by \citet{Gault_1978} shows asymmetry in the ejecta curtain of a 6 km/s $\theta_I = 60^\circ$ impact into quartz sand and Figures 2 and 3 by \citet{Suo_2024} show asymmetry in the ejecta curtains of 104 m/s 
oblique impacts into sand up to and including $\theta_I = 80^\circ$.  
%Also \citet{Schultz_1999} described ejecta curtains   from 1.0-1.5 km/s impacts into sand at $\theta_I = 30^\circ$ as being asymmetric. 
For high velocity impacts (few km/s and higher), and
at grazing impact angles of below $20^\circ$, there can be a forbidden zone where no ejecta lands,  
and giving an ejecta distribution called a butterfly pattern \citep{Gault_1978}. 
 
Normal impacts on a homogeneous, isotropic and level target substrate exhibit 
rotational symmetry about an axis normal to the surface that intersects the impact point. Azimuthal symmetry facilitates integrating over the ejecta mass and velocity distributions,  consequently scaling laws based on dimensionless parameters for estimating crater size and ejecta mass and velocity distributions predominantly apply to normal impacts \citep{Holsapple_1993,Housen_2011,Kuichi_2019,Celik_2022}.

The momentum transfer caused by an impact can be computed from the integral of the fraction of ejecta distribution that escapes the planetesimal (e.g., \citealt{Dobrovolskis_1984,Farinella_1992,Raducan_2022}).
An estimate for the torque caused by impacts on asteroids requires taking into account the distribution of impact angles in the population of projectiles \citep{Dobrovolskis_1984, Luniewski_2024}. 
This motivates us to improve upon estimates for the ejecta distributions of oblique impacts.

Impact craters formed at normal impact angle obey scaling relationships for crater size and volume 
\citep{Holsapple_1993,Housen_2011,Celik_2022,Mazur_2022} 
based on the dimensionless parameters  
\begin{align}
    \pi_2 & = \frac{g a}{ U^2} \\ % \frac{g_a a_{pj}}{ u_{pj}^2} \\
    \pi_3 & = \frac{Y}{\rho U^2} \\ % \frac{Y_a}{\rho_a u_{pj}^2} \\
    \pi_4 & = \frac{\rho}{\delta} .   % \frac{\rho_a}{\rho_{pj}} .
\end{align}
%Here $g_a$ is the gravitational acceleration at the impact site,
%$\rho_a$ is the density of the substrate, $Y_a$ is substrate strength, $u_{pj}$ is projectile velocity, $\rho_{pj}$ is projectile density, and $a_{pj}$ is the projectile radius.   We use subscript $pj$ to refer to projectile properties  and a subscript  $a$ to refer to asteroid or parent body physical or material properties. 
Here $g$ is the gravitational acceleration at the impact site,
$\rho$ is the mean density of the substrate, $Y$ is substrate strength, $U$ is projectile velocity, 
$\delta$ is projectile density, and $a$ is the projectile radius.  
 %We use subscript $pj$ to refer to projectile properties  and a subscript  $a$ to refer to asteroid or parent body physical or material properties. 
Recent compilations of crater dimensions suggest that these scaling laws are remarkably good at matching crater properties, such as radius and volume, over a wide range of impact and substrate properties, 
\citep{Holsapple_1993,Housen_2011,Housen_2018,Mazur_2022}, 
and including a low velocity ($<$ 1 m/s) regime \citep{Kuichi_2019,Celik_2022}. 
Assuming an approximation known as the point-source approximation,  \citet{Housen_2011} developed 
power law scaling relations for the ejecta curtains of normal impacts, with 
 ejecta properties such as ejecta speed and ejected mass dependent on the horizontal distance from the impact point.   
 
Ejecta mass and velocity distribution functions for oblique impacts, when integrated,  are directly related to the momentum transfer efficiency and the change in spin caused by an impact.   
Simulations of high velocity oblique impacts have found that ejecta distributions are sensitive to azimuthal angle \citep{Luo_2022,Raducan_2022}.  For grazing impacts, 
 ejecta thickness uprange (toward the projectile launcher) can be low due to reduced ejecta launch angles and velocities giving the butterfly pattern \citep{Luo_2022}. 
The ejecta distributions derived in the point-source approximation by \citet{Housen_2011} were modified to depend upon azimuthal angle and ejecta angle and then fit to velocity and mass ejecta distributions measured from shock physics code simulations \citep{Raducan_2022}.  
 %as a function of impact angle and azimuthal angle
%
\citet{Dobrovolskis_1984} used a power law form for the ejecta distributions, based on $\sim 6$ km/s normal  impact experiments by \citet{Stoffler_1975}, to estimate a cumulative torque on asteroids due to impacts. Improved measurements of ejecta distributions for oblique impacts would help improve estimates for momentum transfer and torque caused by impacts (e.g., \citealt{Luniewski_2024}).

Ejecta from laboratory high velocity (1 to 6.5 km/s) oblique impacts have been tracked using high speed cameras \citep{Anderson_2003,Anderson_2004,Anderson_2006}
giving measurements of ejecta angle and velocity distributions for different impact angles $\theta_I$ and as a function of azimuthal angle $\zeta$ measured in the substrate plane.  
Azimuthal variations in ejecta angle and velocity imply that there is asymmetry in the subsurface response  \citep{Anderson_2003,Anderson_2004,Anderson_2006} and 
this was verified for approximately 100 m/s oblique impacts into sand using embedded accelerometers \citep{Suo_2024}.  
Simulations of high velocity impacts have been used to measure the sensitivity of ejecta velocity, angle and mass distributions to both impact and azimuthal angle \citep{Raducan_2022,Luo_2022}. 

%{\bf goals of paper here:}
Ejecta particle velocity can be tracked in videos taken with high speed cameras \citep{Anderson_2003,Anderson_2004,Anderson_2006,Tsujido_2015}, but because particles in an ejecta curtain can obscure material behind them, it is more challenging to measure the ejecta mass distribution from 
videos. 
Trays laid on the surface or mounted above the surface can be used to capture ejecta which  
 later are weighed, giving constraints on the mass distribution of the ejecta curtain  \citep{Stoffler_1975,Wunnemann_2016,Luther_2018,Mazur_2022}.  
In this study, 
we build upon our prior experimental study of intermediate velocity ($\sim 100$ m/s) oblique impacts
 \citep{Suo_2024} by similarly using trays to catch ejecta at different azimuthal angles. 
 % Measurements of the mass  distribution of ejecta curtains has predominantly focused on normal impacts. 
%In this paper we aim to describe experimental measurements of the mass in ejecta for oblique impacts   at an intermediate velocity into a granular substrate, sand.   
We use crater topographic profiles at different impact angles to model the azimuthal and impact angular dependence of crater radius. 
We extend ejecta crater scaling models for normal impacts 
to describe the azimuthal and impact angle dependence of the ejecta mass distribution.   
Our estimated distributions may be used to guide models that integrate over ejecta properties to estimate momentum transfer and torque from impacts at speeds near 100 m/s. 

While most impacts in the inner solar system occur at high velocities, with a mean of about 5 km/s \citep{Bottke_1994}, lower velocity impacts occur due to ejecta that form secondary craters, 
in the outer solar system, and on planetesimals while they are embedded in a disk.   
A low or intermediate velocity regime (of about 20 to 100 m/s) is relevant for impacts onto planetesimals by particles in the protosolar nebula, leading to accretion or causing erosion,  \citep{Quillen_2024}.
In the outer solar system, most impacts are below 4 km/s \citep{Greenstreet_2019}.  
Approximately 65\% of impacts onto Classical trans-Neptunian object 486958 Arrokoth are estimated to originate from a cold classical Kuiper belt population of objects 
(see Table 1 by \citealt{Greenstreet_2019}) and with impact velocity distribution
 peaking at about 300 m/s (see Figure 1 by 
\citealt{Greenstreet_2019}).   In the inner solar system, 
particles ejected during a high velocity impact can later 
hit a moon or a planet's surface at lower velocity forming a secondary crater (e.g., \citealt{Bierhaus_2012}).
Studies of low and intermediate impacts are also relevant for the design of landers (e.g., \citealt{Ho_2021}).   
 
Crater scaling laws (e.g., 
\citealt{Holsapple_1993,Housen_2011,Housen_2018,Mazur_2022,Celik_2022,Neiderbach_2023})  are remarkably successful at estimating numerous physical quantities including crater depth, volume, radius, rim uplift, and ejecta velocity and mass distributions over a wide range in parameters such as projectile mass,  impact velocity, gravitational acceleration and material properties.  Physical quantities (such as crater volume) are estimated with power laws which are functions that do not separate or delineate between low and high velocity impact regimes.   
In the study present here, we characterize the angular distribution of ejecta from $\sim $ 100 m/s oblique impacts. 
Subsurface seismic pulses for $\sim $ 100 m/s impacts in sand exhibit phases analogous to those characteristic of high velocity impacts \citep{Suo_2024}:
 compressive, pressure release and excavation phases, even though shocks are not present. 
 Despite the successes of dimensionless crater scaling laws, 
 care should be taken if and when applying the results of our study to interpret impacts at higher or lower impact velocities than those of our experiments. 

%In new experiments  we use trays to  measure ejecta mass in bins of different azimuthal angles.    

\section{Experiments}

\begin{figure*}
    \centering
    \includegraphics[width=3truein]{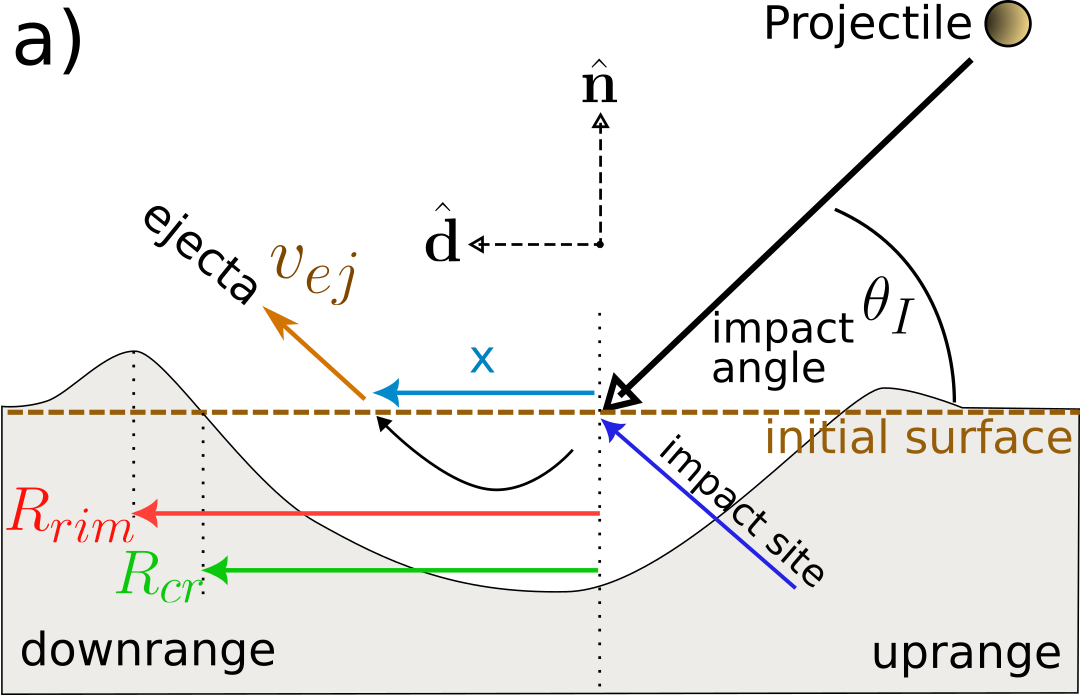} % fig1a
    \includegraphics[width=2.2truein]{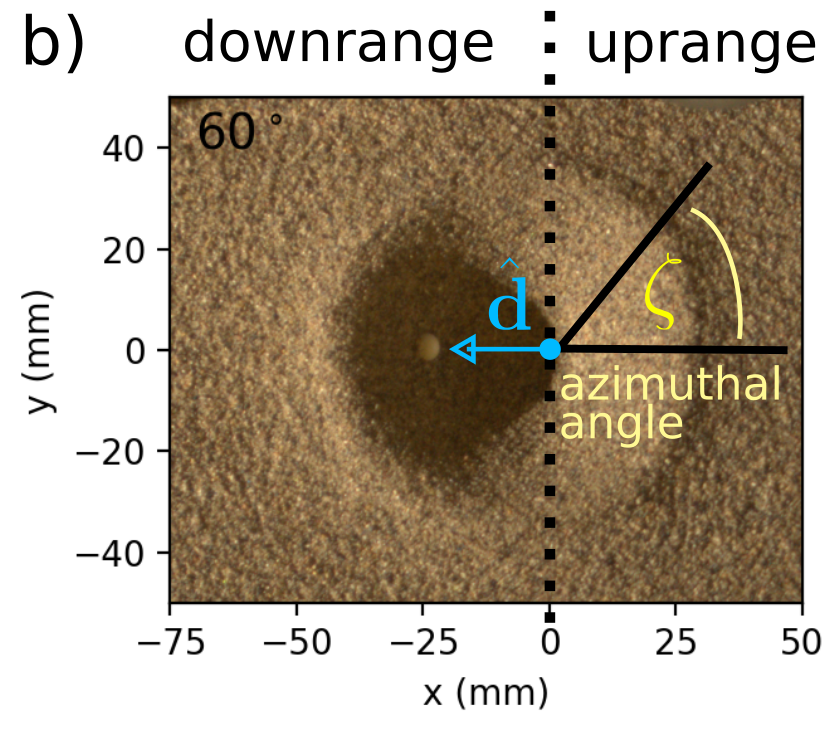} %fig1b
    \caption{a) We show a side view illustration of a transient crater created by an oblique impact at impact angle $\theta_I$.  Distances are measured in the plane of the level surface prior to impact. The crater radius  $R_{cr,slv}$  is the distance between impact point and where the crater surface crosses the level of the surface prior to impact (following \citealt{Housen_2011}). The  crater rim radius  $R_{rim}$ is the distance in the surface level plane between 
    impact point and rim peak.  The ejecta launch point is given in terms of $x$ which is a distance from the  impact point.  We also illustrate 
    normal and downrange directions $\hat {\bf n}$ and $\hat {\bf d}$. 
    The tan shading shows the transient crater. The brown dashed line shows the level of the surface prior to impact. 
    b) A view from above illustrating the downrange direction and the azimuthal angle $\zeta$ which is measured with origin at the impact point. With projectile originating from the right, downrange is to the left. The background is a photograph of a laboratory transient crater with an impact angle of $\theta_I  = 60^\circ$.  After impact, the BB projectile remained in the crater and can be seen on the left.  The scale is in mm.  The photograph is the same as one shown by \citet{Suo_2024}. }
    \label{fig:crater_dims}
\end{figure*}

Our experiments build upon those described by \citet{Suo_2024}, with  
projectile and target substrate properties summarized in Table \ref{tab:airsoft_bbs}.   
Illustrations of our experiments are shown in Figure \ref{fig:crater_dims}. 
New impact experiments presented here are carried out with the same airsoft gun, the same projectiles, 
into the same tub containing the same sand. 
%These are the same values as used in  the experiments by  \citep{Suo_2024}. 
%The projectile and substrate properties are similar to those described by  \citep{Suo_2024}. 
Small plastic spherical projectiles (referred to as pellets or BBs) are launched with an airsoft gun\footnote{An airsoft gun is a low-power smoothbore air gun that usually shoots small plastic spheres.} at a speed of $U = 103$ to 105 m/s.   We chose an 
airsoft gun to launch projectiles because it is low cost, it is safer than a projectile launcher that fires high velocity projectiles (important as our lab in located on a university campus), it fires in an intermediate velocity regime, and the density of the plastic BBs that the gun fires is similar to our target substrate density and so  representative of most astronomical impacts.   

The target substrate is fine sand with grain semi-major axis mean value $ \approx 0.3$~mm, as described in previous experiments  \citep{Wright_2020b}.
Dimensionless constants $\pi_2, \pi_3, \pi_4$, used to characterize crater scaling regimes,  
are also listed in Table \ref{tab:airsoft_bbs} along with their definitions (following \citealt{Holsapple_1993,Housen_2011,Housen_2018}).  
Experimental measurements suggest that rocky granular or regolith systems would have an effective strength of order $Y \approx 500$ Pa \citep{Brisset_2022} and we adopt that value to compute  
the dimensionless parameter $\pi_3$. 

\begin{table}[ht]
    \centering
    \caption{Properties of the projectiles and target substrate and dimensionless numbers}
        \label{tab:airsoft_bbs}
    \begin{tabular}{lll}
    \hline
    Quantity & Symbol & Value \\
    \hline
     Projectile Mass     & $m$ & $0.20 \pm  0.002$ g \\
     Projectile Radius  &  $a$ & $2.98 \pm 0.005$ mm \\  %   Projectile Radius  &  $a_{pj}$ & $2.98 \pm 0.005$ mm \\
     Projectile Density  & $\delta$ & 1.80 g cm$^{-3}$ \\ %    Projectile Density  & $\rho_{pj}$ & 1.80 g cm$^{-3}$ \\
    Projectile Speed   & $U$ & 104 $\pm 1 $ m/s \\  %   Projectile Speed   & $u_{pj}$ & 104 $\pm 1 $ m/s \\
    % Composition & \multicolumn{2}{l}{ Polylactic acid (PLA plastic)} \\    
    %Substrate & sand & \\
    %Sand grain size & $a_s$ & 0.3 mm \\
    Target  density & $\rho$ & 1.5 g cm$^{-3}$ \\ %    Target  density & $\rho_a$ & 1.5 g cm$^{-3}$  \\ % sand
    Washtub rim radius\!\! & $R_{\rm tub}$ & 25.1 cm\\
    Washtub depth & $H_{\rm tub}$ & 25 cm\\
   % Crater diameter & $D_{rim,n}$ & 6.8 cm \\
    Crater radius & $R_{cr,slv,n}$ & 2.85 cm \\
    Crater volume & $V_{cr,slv,n}$ & 13.84 cm$^{-3}$ \\
    \hline 
    Dimensionless params.\!\! & Formula &  Value \\
    \hline
    $\pi_2$ & $g a_/U^2$ & $2.7\times 10^{-6}$ \\ %    $\pi_2$ & $g_a a_{pj}/u_{pj}^2$ & $2.7\times 10^{-6}$ \\
    %$Fr$ & $u_{pj}/\sqrt{g_a a_{pj}}$ & 608 \\
   $\pi_3$  & $Y/(\rho U^2)$ & $3 \times 10^{-5}$ \\ %    $\pi_3$  & $Y_a/(\rho_a u_{pj}^2)$ & $3 \times 10^{-5}$ \\
    $\pi_4$ & $\rho/\delta$ & 0.83  \\  % $\pi_4$ & $\rho_a/\rho_{pj}$ & 0.83  \\
    $\pi_R$ & $R_{cr,slv,n}\! \left({\rho/}{m}\!\right)^\frac{1}{3}$\!\! & $5.6$\\
%    $\pi_R$ & $R_{cr,slv,n}\!\! \left({\rho_a/}{m_{pj}}\!\right)^\frac{1}{3}$\!\! & $5.6$\\
    $\pi_V$ & $V_{cr,slv,n} \rho/m$ & $110$ \\
%    $\pi_V$ & $V_{cr,slv,n} \rho_{a}/m_{pj}$ & $110$ \\
    %     $\tau_{ex}$ & $\sqrt{R_{cr,slv}/g}$ & 60 ms \\
        % & & 0.75 g cm$^{-3}$ millet \\
     \hline
    \end{tabular}
    \begin{tablenotes}  \small \item
    The parameter $\pi_3$ is computed with  $Y$=500 Pa.  The crater radius $R_{cr,slv,n}$ and
    volume $V_{cr,slv,n}$ were measured in our experiments, and are not based on scaling estimates. 
    \end{tablenotes}
\end{table}
%{\textbf{Add errors on the quantities in the table \ref{tab:airsoft_bbs}}.
% note PLA is supposed to have a density of 1.2 to 1.4 g/cc, whereas ABS has an even lower density of 1.06 to 1.08 g/cc.  Our estimated density is lower than typical of PLA?
% ratio of crater rim to zero radius is 1.2 which is a bit lower than 1.3 given by \citet{Housen_2011} as typical for cohesionless granular materials

%As shown in Figure \ref{fig:crater_dims},  the distance between impact point and crater rim peak we denote $R_{rim}$.  
%The crater radius $R_{cr,slv}$ is measured from the impact point to the location where the crater depth profile radius crosses the level of the surface prior to impact, as shown in  Figure \ref{fig:crater_dims}a and following \citet{Housen_2011}. 

Following \citet{Housen_2011}, a particle is considered ejected as it crosses the plane that was the 
surface of the target substrate prior to impact, as shown in Figure \ref{fig:crater_dims}a.  In this plane, 
the distance from impact is $x$ and $v_{ej}$ is the ejection velocity. After the transient crater has formed, we take crater radius $R_{cr,slv}$ to be the radius from the point of impact where the crater surface crosses the level of the surface prior to impact, and as shown in Figure \ref{fig:crater_dims}a.   
The subscript $slv$ stands for surface level. The radius $R_{rim}$ is the 
distance from impact point to the rim peak, also measured in the target substrate level plane. 
For oblique impacts we use the same definitions but these two radii are functions of 
impact angle and azimuthal angle; $R_{rim}(\theta_I,\zeta)$ and $R_{cr,slv}(\theta_I,\zeta)$.  

The crater volume $V_{cr,slv}$ is the volume of the crater that lies below the surface level prior to impact. 
For normal impacts, we use a subscript $n$, giving crater radius $R_{cr,slv,n}$, 
crater rim radius $R_{rim,n}$, 
and crater volume $V_{cr,slv,n}$.   Crater radii and volume for normal impacts 
along with dimensionless parameters  $\pi_R, \pi_V$ computed from these quantities are also 
listed in Table \ref{tab:airsoft_bbs}. 

Crater scaling laws predict that a normal impact in the strength regime has crater radius 
\begin{align}
R_{cr,slv,n}(\pi_3,\pi_4) = a \left(\frac{4\pi}{3} \right)^\frac{1}{3} H_2 
\pi_3^{-\frac{\mu}{2}} \pi_4^{-\nu}, 
\label{eqn:Rcr_str}
\end{align} and in the gravity regime
\begin{align}
R_{cr,slv,n}(\pi_2,\pi_4)  = a \left(\frac{4\pi}{3} \right)^\frac{1}{3} H_1  \pi_2^{-\frac{\mu}{2+\mu}} 
\pi_4^{-\frac{2\nu}{2+\mu}} , \label{eqn:Rcr_grav}
\end{align}
following Table 1 by \citet{Housen_2011}, but with $n$ subscripts to denote a normal impact. 
The parameters $H_1,H_2$ and exponents $\mu,\nu$ are dimensionless but can depend upon the material properties and are estimated empirically using experimental data  \citep{Housen_2011,Housen_2018}. 

A crater is in the strength regime if  $\pi_3^{1+ \mu/2} \pi_4^\nu/\pi_2 > 1$  \citep{Holsapple_1993}.
 % their eqn 18 and rechecked  again!
With exponents $\mu,\nu \sim 0.4$ typical of granular systems  \citep{Housen_2011} and $Y = 500 $ Pa, 
 our experiments have $\pi_3^{1.2} \pi_4^{0.4}/\pi_2 \sim 1 $,  similar to unity.  Our experiments lie near the line dividing strength and gravity regimes.   Because we work in a granular system and 
 its effective strength has not been measured, we often assume that we are in the gravity regime, following 
 %There is some ambiguity on the use of a strength parameter in a granular material.  
\citet{Housen_2011}, who suggested that impacts into granular systems would always be in the gravity regime.  
However, \citet{Scheeres_2010} argued that even a low level of strength would put smaller and lower velocity impactors in a low gravity environment in the strength regime.  
%Our craters are in near the boundary of strength and gravity regimes \citep{Suo_2024}. 
Setting equation \ref{eqn:Rcr_str} or \ref{eqn:Rcr_grav} equal to our normal impact crater radius and using values for the dimensionless parameters from Table \ref{tab:airsoft_bbs}, 
we estimate $H_1 \approx 0.7$ in the gravity regime or $H_2 \approx 0.7$ in the strength
regime.   These coefficients are similar to those estimated 
for different experiments (see  Table 3 by \citealt{Housen_2011}).

We describe oblique impacts with impact angle $\theta_I$ which is the angle between the projectile velocity vector and the target substrate plane.  The impact angle is zero if the impact is grazing and is 
$\pi/2$ if the impact is normal to the surface, as shown in Figure \ref{fig:crater_dims}a.    
The azimuthal angle $\zeta$ is measured 
using the point of impact as origin and in the plane perpendicular to the surface normal $\hat {\bf n}$ as shown in Figure 
\ref{fig:crater_dims}b.  We adopt the convention that $\zeta = 0$ in the uprange direction (opposite the $\hat {\bf d}$ vector), and toward the projectile launcher, following \citet{Anderson_2004,Raducan_2022}. 
%The goal of these experiments are to study the mass distribution of impact ejecta as a function of azimuthal angle, resulting from an oblique impact.  Here 

\subsection{Ejecta mass at different azimuthal angles}
\label{sec:ejmass}

\begin{figure}[ht!]\centering
\includegraphics[width = 3 truein]{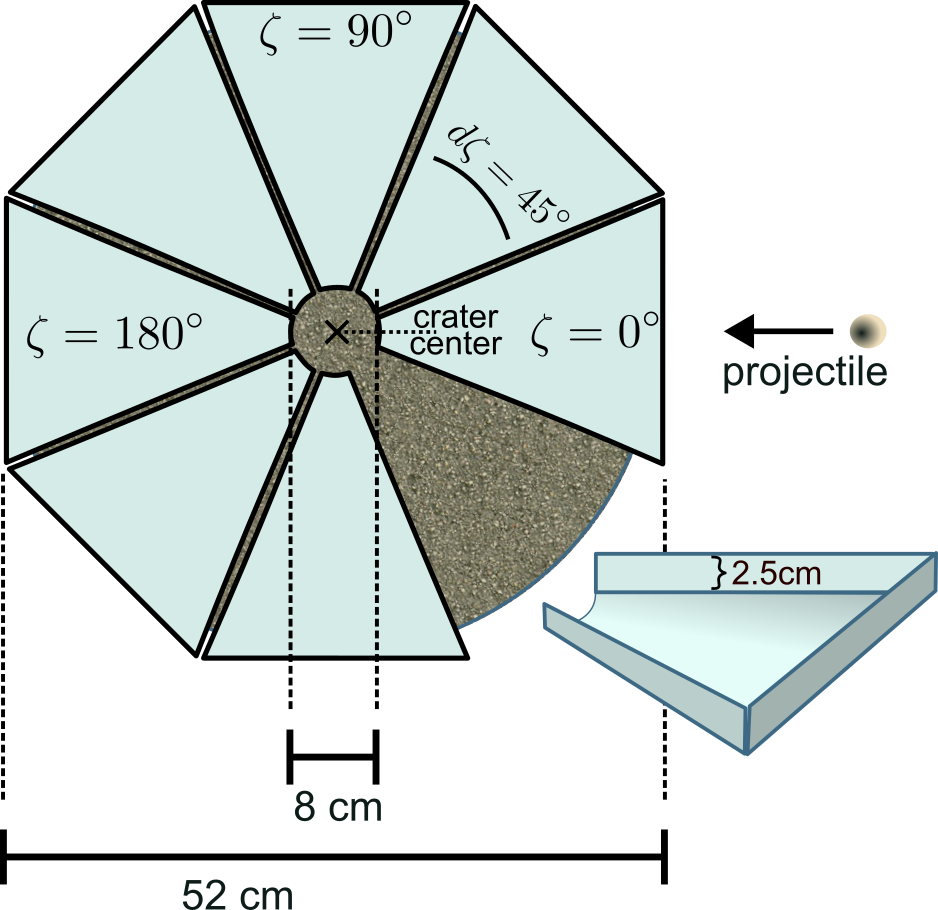}  %fig2
\caption{Orientation and shape of eight paper trays used to catch ejecta mass as seen from above. 
The projectile originates from the right.   The crater center is centered in the central hole created by
the paper trays. One of the paper trays is shown on the lower right with orientation chosen to show its 
folded edges. 
  \label{fig:trays}}
\end{figure}

\begin{figure}[ht!]\centering
\includegraphics[width = 3 truein]{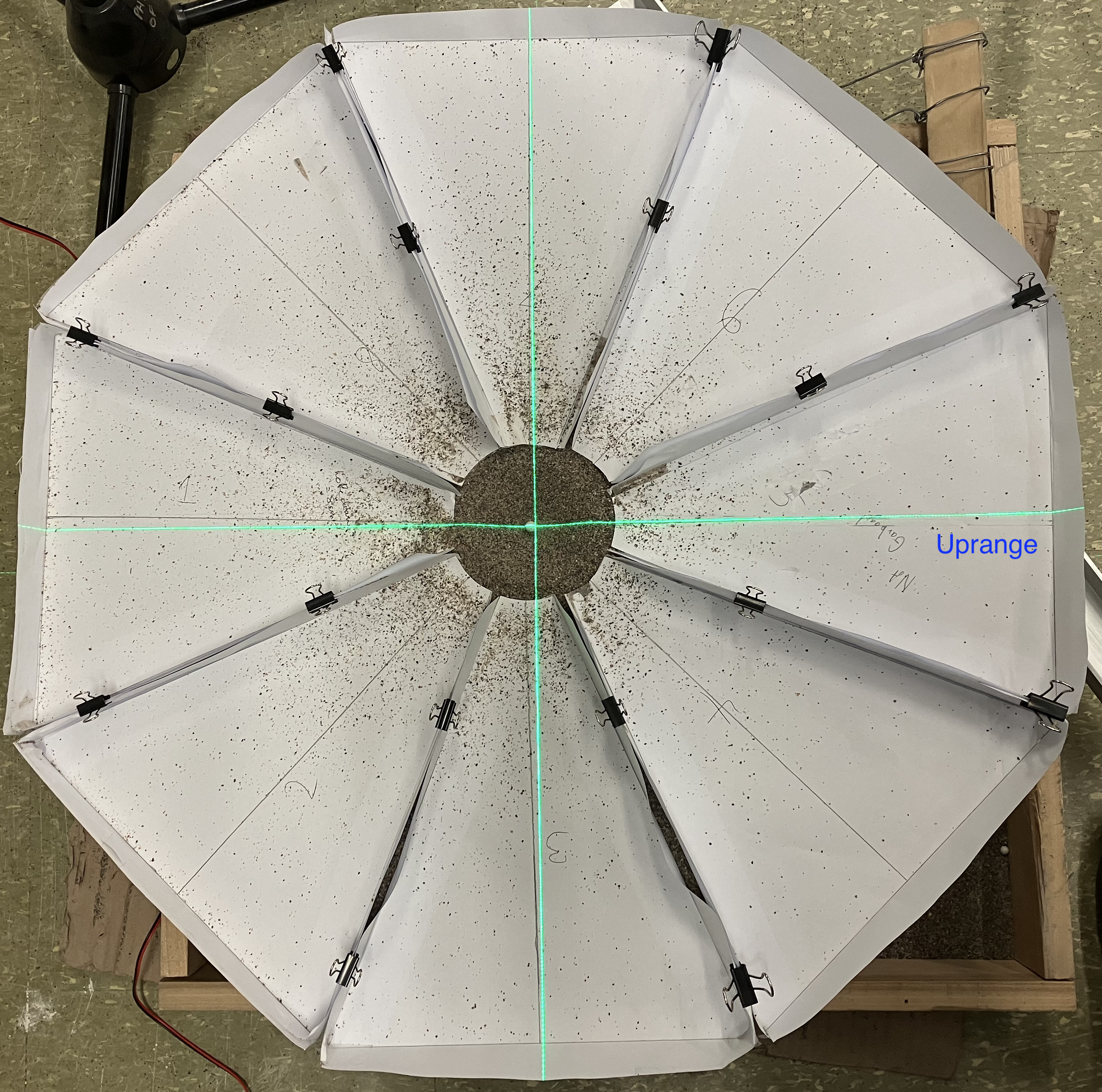} % fig3
\caption{A photograph of the trays covered in sand ejecta after an impact at $\theta_I=60^\circ$. 
The photo was taken from above.     The green laser cross helps us 
aim the projectile and center the trays before impact.  The projectile came from the right side and there
is more ejecta in the downrange bins on the left.  
 \label{fig:photo}}
\end{figure}

\begin{table*}[ht]
    \centering
    \caption{Ejecta mass $M_{ej}(\theta_I,\zeta)d\zeta$ in g measured as a function of impact and azimuthal angles}
    \label{tab:mej}
\begin{tabular}{lllllllllllll}
\hline
$\theta_I/\zeta$ & 0 & 45 & 90 & 135 & 180 & 225 & 270 & 315  \\
\hline
20& 0.10$\pm$0.14& 0.16$\pm$0.18& 0.32$\pm$0.33& 0.35$\pm$0.22& 0.26$\pm$0.21& 0.45$\pm$0.29& 0.41$\pm$0.29& 0.20$\pm$0.29 \\
%20& 0.04$\pm$0.08& 0.09$\pm$0.15& 0.19$\pm$0.19& 0.33$\pm$0.23& 0.27$\pm$0.21& 0.43$\pm$0.31& 0.24$\pm$0.19& 0.01$\pm$0.00 \\
30& 0.13$\pm$0.08& 0.23$\pm$0.13& 0.47$\pm$0.13& 0.70$\pm$0.17& 0.85$\pm$0.27& 0.64$\pm$0.19& 0.46$\pm$0.14& 0.33$\pm$0.15 \\
40& 0.05$\pm$0.05& 0.19$\pm$0.11& 0.65$\pm$0.14& 1.37$\pm$0.21& 1.81$\pm$0.23& 1.49$\pm$0.31& 1.05$\pm$0.31& 0.28$\pm$0.12 \\
50& 0.23$\pm$0.18& 0.35$\pm$0.18& 0.63$\pm$0.25& 1.24$\pm$0.27& 1.05$\pm$0.19& 1.76$\pm$0.38& 0.95$\pm$0.30& 0.35$\pm$0.12 \\
60& 0.24$\pm$0.19& 0.34$\pm$0.22& 0.60$\pm$0.15& 1.34$\pm$0.32& 1.67$\pm$0.54& 1.35$\pm$0.45& 0.67$\pm$0.20& 0.42$\pm$0.32 \\
70& 0.30$\pm$0.07& 0.40$\pm$0.15& 0.76$\pm$0.20& 1.22$\pm$0.11& 1.72$\pm$0.26& 1.61$\pm$0.25& 0.93$\pm$0.30& 0.40$\pm$0.12 \\
80& 0.40$\pm$0.13& 0.65$\pm$0.33& 0.96$\pm$0.39& 1.34$\pm$0.40& 1.40$\pm$0.34& 0.98$\pm$0.17& 0.78$\pm$0.26& 0.52$\pm$0.07 \\
90& 0.87$\pm$0.26& 0.74$\pm$0.13& 0.84$\pm$0.16& 0.72$\pm$0.19& 0.81$\pm$0.18& 0.84$\pm$0.17& 0.89$\pm$0.18& 0.95$\pm$0.18 \\
\hline
\end{tabular}
\begin{tablenotes}  
      \small \item
Each entry shows the ejecta mass in g as a function of azimuthal angle $\zeta$ and
impact angle $\theta_I$  in trays of angular width $d\zeta = 45^\circ$ as  shown 
in Figure \ref{fig:trays}. 
Each row shows a single impact angle $\theta_I$ which is given in the leftmost column in degrees. 
The columns are headed by azimuthal angle $\zeta$ in degrees. 
Each entry gives the average of measurements from 6 experiments.
The uncertainties show the standard deviation computed from the scatter of these measurements. 
%The top row shows the central azimuthal angle for each tray.  
An angle of $\zeta =0$  corresponds to uprange (toward the projectile launcher).  
\end{tablenotes}
\end{table*}

We carried out a new set of oblique impact experiments at different impact angles specifically to measure ejecta mass. 
To collect ejecta from our laboratory impacts, we constructed a set of eight nearly identical paper trays to catch ejecta.  When placed next to each other, they cover an octagonal area around the impact point, as shown in Figure \ref{fig:trays} and in the photo in Figure \ref{fig:photo}.  Paper was used to keep the trays light weight, reducing compression of the sand target substrate from the weight of the tray. Each paper tray weighs about 5.25 g.  
A circular area 8 cm in diameter, approximately matching crater diameter, was cut out from the apexes of the trays to allow ejecta to escape the impact region.  Each tray covers an azimuthal angle of $d\zeta = 45^{\circ}$, as illustrated in Figure \ref{fig:trays}.  
Each tray is labelled by its central azimuthal angle.   

More ejecta mass originates at lower velocity near the crater rim than at higher velocity 
nearer the impact point (e.g., \citealt{Housen_1983}). 
If a tray overlaps the crater rim, it would block the ejecta curtain.  
To capture the most ejecta mass, we designed the trays
so that the tray inner edges are close to the crater rim, but do not cover it.  
 In the mid-velocity regime of our 104 m/s impacts, craters are nearly round, 
with larger major to minor axis ratio at lower impact angles 
(at $\theta_I = 20^\circ$, the major to minor crater axis ratio is only 1.18; see Table 5 by \citealt{Suo_2024}).
As a consequence, we cut the tray inner edges so that they are arcs of a circle.   The three straight
sides of the trays have folded edges, as shown on the lower right in Figure \ref{fig:trays} so that ejecta
does not bounce out just after impact or fall out when we move the tray to weigh it. 
The inner central curved tray edge is flat so that ejecta is not blocked by the tray.  
We aim the projectile so that the resulting crater is centered with respect to the central 
region of sand that is not covered by the trays, as illustrated in Figure \ref{fig:trays}.
We did not use smaller area trays because the ejecta mass collected in each tray is low, 
often less than a gram.  

\begin{figure}[htbp]\centering
\includegraphics[width=3truein]{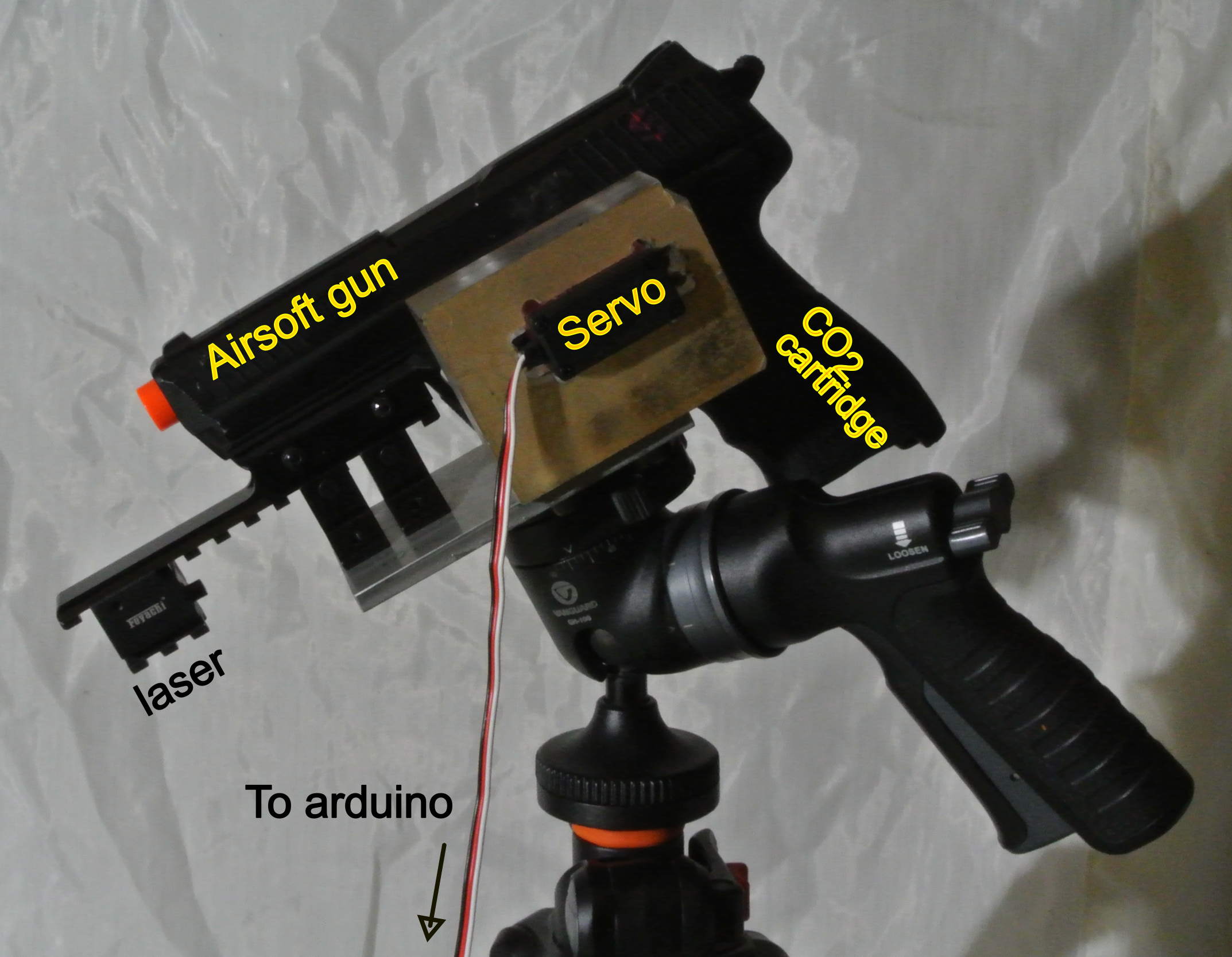} % fig4
\caption{An annotated photograph of the airsoft gun we use to launch our BB projectiles. 
\label{fig:airsoft}
}
\end{figure}

To improve projectile aim, 
we rebuilt the airsoft gun mount and digital servo trigger system so that it is held 
more securely than in our previous experiments (see Figure \ref{fig:airsoft}).  The airsoft gun is fired electronically,  
without touching it.  This was also true in the experiments by \citet{Suo_2024} but then the air-soft gun was mounted to a piece of plywood which we subsequently replaced with a piece of machined aluminum. 
We also mounted an adjustable laser sight to an extension of the airsoft gun's Picatinny rail\footnote{The Picatinny rail, also known as the MIL-STD-1913 rail, is an American standard system that provides a mounting platform for firearm accessories.  The rail consists of flattened T coupled with a hexagonal top.  Accessories such as lasers sights, are slid into place along the T and locked into place by tightening the hexagonal top. }.  which  helps us aim the projectile and measure the impact angle. 
Prior to carrying out experiments with the trays, we do a set of impacts to adjust the airsoft gun tripod angle and location so that the crater is centered
in the washtub container.  
The impact angle is measured with the laser sight and a large protractor that is drawn 
on a sheet of poster board, and is accurate to approximately 2 degrees. 
%After we have set the crater location and impact angle, we run series of experiments with the paper trays to catch ejecta. 

Prior to each impact experiment, the surface is raked with a wire rake with wire tongs that are about 10 cm deep,  as described in our prior experiments \citep{Suo_2024}.  After raking, 
 the surface is leveled with a rectangular metal bar so that its level is equal to the height of the rim of the tub holding the sand. Afterward raking and leveling, the empty paper trays are placed on the surface 
with positions and orientations illustrated in Figure \ref{fig:trays}. 
%For each impact angle we adjusted the impact laser target so that the crater would be  centered within the hole made by the trays. 
% For oblique impacts, the impact site is uprange of the crater center (measurements of the distance between crater center and impact site are given by \citealt{Suo_2024}).
After an impact we examine the crater.  A few experiments were discarded because the crater rim 
did not lie within the central hole made by the trays.  Otherwise,  
after each impact, the weight of the sand collected in each tray is measured with a digital scale. 

We discuss uncertainties in our ejecta mass measurements. 
The distance from impact point to the center of the tray pattern 
gives a variation in the azimuthal angle (measured with origin at the impact point) subtended 
by each tray.   
Despite making improvements in our trigger servo system, we found that the impact point 
is not exactly at the same position every time we fire a projectile.  By examining 
crater locations with respect to a laser target, we find that the difference between desired
and actual impact point has a standard deviation of about a cm. 
Ejecta scaling is done with origin at the impact point \citep{Housen_2011}, as illustrated in Figure \ref{fig:crater_dims}a, however, we attempted to center our trays 
with respect to the crater center, as illustrated in Figure \ref{fig:trays}.   
Craters from oblique impacts are not centered about the impact point.  
To characterize the distance between 
crater center and impact point,  \citet{Suo_2024} measured 
 the distance $d_{ai}$ between the midpoint of uprange and downrange rim peaks and the point of impact.  The distance $d_{ai}$  is a maximum of 1.49 cm
at impact angle $\theta_I = 40^\circ$, and below 1.1 cm for the remaining impact angles, 
 as listed in Table 5 by \citet{Suo_2024}.     The crater at $\theta_I=40^\circ$ may be particularly elongated
in part because the projectile passed through the ejecta curtain. 

We consider the value  $d_{ai} = 1.1$ cm (a value characteristic of our experiments for $\theta_I$ between 30 and 60$^\circ$), 
the angle subtended by the $\zeta = 180^\circ$ downrange tray at its inner edge is about 0.8 times $d\zeta = 45^\circ$ and the angle subtended by the 
$\zeta = 0$ uprange tray is 1.4 times $d\zeta = 45^\circ$.   This implies that our ejecta masses
are lower than they should be
 in the downrange trays and higher than they should be in the uprange trays. 
A similar size variation in angle subtended by the trays is caused when the impact point 
is 1~cm closer to the $\zeta=90^\circ$ tray than the $270^\circ$ tray, in which case
the trays on the closer side would  
receive more ejecta than the opposite side.   
The worst case error in subtended angle of about 40\% seems large,  however we find that 
the azimuthal variations in 
ejecta mass are so large that they dwarf uncertainties caused by errors 
in impact point with respect to the trays. 
% The error caused by tray centering is largest for the midrange of impact angles $\theta_I$ between 30 and 60$^\circ$ where $d_{ai}$ is largest, with error possibly as large as 40\%. 
%The downrange and uprange correction  factors of 0.8 and 1.4 can be considered the worst case as much of the ejecta does not
% land right next to the tray edge, and $d_{ai}$ is particularly large at $\theta_I = 30$ to  $50^\circ$ in part 
% because the projectile ricochets, passing through ejecta curtain or crater rim edge.  

Because craters are smaller and more elongated at grazing angles, with $\theta \le 30^\circ$ the trays would miss 
some ejecta near the crater rim, particularly along $\zeta \sim 90^\circ$ and $270^\circ$. 
However, variations in ejecta mass caused by crater elongation are much smaller than those caused by craters that are not centered with respect to the trays.   An offset between crater center and trays 
puts more ejecta on one side than the other.  As the craters are larger and closer to the tray edges 
at higher impact angle, an asymmetry in the ejecta mass could be overestimated in the higher impact angle 
experiments compared to the lower impact angle experiments. 

High velocity ejecta can land past the outer edge of our trays.   For a few experiments we placed a large cardboard catcher with a high backside behind the washtub on the downrange side to obtain any high velocity ejecta that escaped our other paper trays.  The catcher covered an azimuthal range of about $120^\circ$. 
At an impact angle $\theta_I = 40^\circ$ we measured the fast ejecta mass that escaped  our trays altogether, 
finding that it is at most 0.16 g and negligible.  The low fraction of high velocity ejecta that we fail to catch with our paper trays 
 is consistent with a low fraction of ejecta mass predicted at high ejecta velocity in normal impacts (from equation 18  by \citealt{Housen_2011}).  
 At $\theta_I=30^\circ$ we measured an escaping ejecta mass of 0.35 g which is less than half
the size of ejecta mass in the uprange tray, and less than 1/10 of the total ejecta mass. 
These two measurements suggest that the 
 fraction of downrange high velocity ejecta that escapes increases with decreasing impact angle. 
 
In some of the experiments at a low impact angle of $\theta_I = 20^\circ$ we measured no or little ejecta 
in the uprange direction.  The high speed imaging of the ejecta curtains shown in Figure 2 by \citet{Suo_2024} shows that ejecta is expelled in the uprange direction  in the $\theta_I = 20^\circ$ impacts 
so a zone of avoidance that is characteristic of a butterfly ejecta blanket might be absent at this 
impact angle.   However, the snap shots from the high speed imaging don't clearly show
ejecta expelled in the uprange direction for the $\theta_I = 10^\circ$ impact.  It is possible that 
the ejecta blanket resembles a butterfly pattern at grazing impact angles, similar 
to the butterfly patterns seen in simulations \citep{Luo_2022} and experiments \citep{Gault_1978} of high  velocity (few km/s) impacts.  Future experiments would be be required to follow up this possiblity. 
 
\begin{figure}[ht!]
    \centering
    \includegraphics[height=7.5truein,trim = 8 18 18 0, clip]{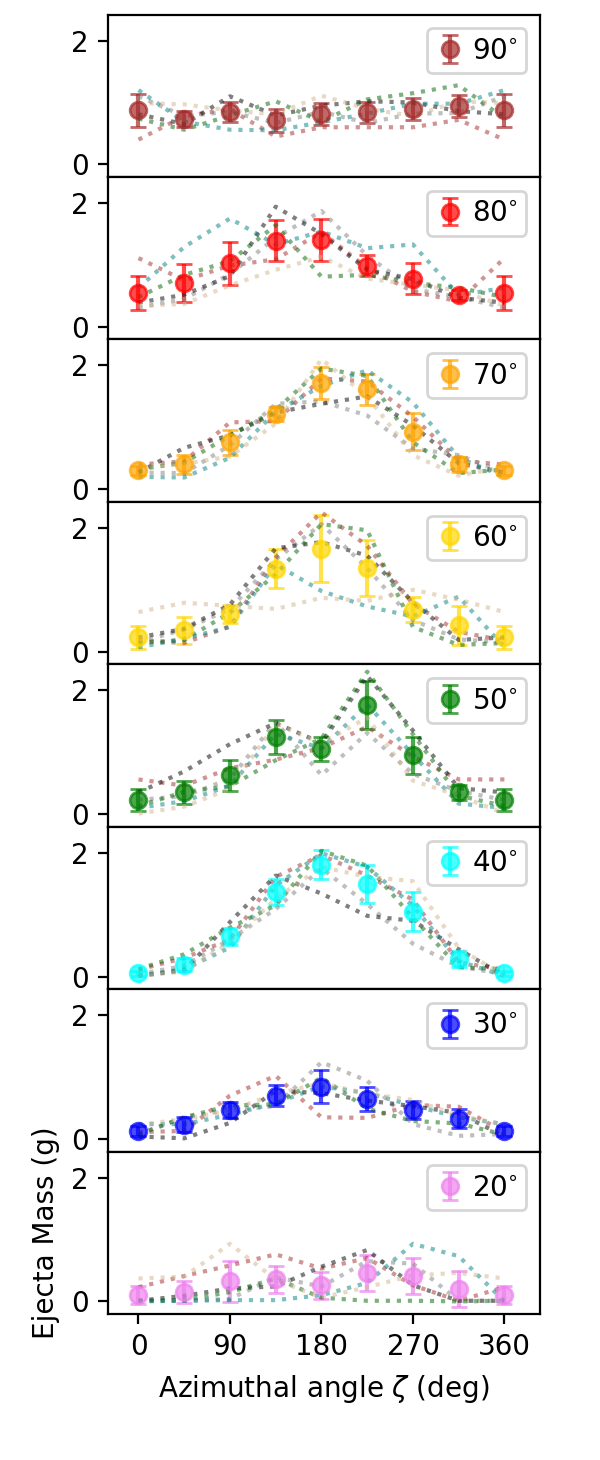} % fig5
    \caption{Ejecta mass in g per $d\zeta = 45^\circ$ azimuthal bin measured at different impact angles. 
    Dotted lines show individual experiments whereas the large dots show averages and vertical error bars show  standard deviations computed from the scatter from 6 different experiments at each azimuthal angle. %The horizontal error bars show the tray angular widths.
    }
    \label{fig:aves}
\end{figure}

We carried out 6 experiments at each of 8 impact angles ranging from 20 to $90^\circ$ with an increment of $10^\circ$.  The average ejecta masses in g (of 6 measurements) as a function impact angle and tray central azimuthal angle are listed in Table \ref{tab:mej}.   
Each panel in Figure \ref{fig:aves} shows the ejecta masses in grams as a function 
of azimuthal angle $\zeta$ (along the horizontal axis) at a different 
impact angle. 
% In Figure \ref{fig:aves},  the dotted lines show individual measurements from each experiment. The large dots at each azimuthal angle show the average ejecta masses computed from 6 experiments.  The vertical error bars  show the standard deviations computed from the scatter about the average values and are also listed in Table \ref{tab:mej}.  

%The horizontal error bars in Figure \ref{fig:aves} show the $d \zeta = 45^\circ$ angle subtended by each tray.  

Figure \ref{fig:aves} shows that ejecta mass from oblique impacts depends on azimuthal angle,  
with more mass ejected in the downrange direction than in the uprange direction.  
At $\theta_I=50^\circ$,
there is a dip in the ejecta mass  at azimuthal angle $\zeta = 180^\circ$, 
in the down range direction.  At this impact angle, the projectile ricochets at a low
angle, grazing the crater rim (see Figure 4 by \citealt{Suo_2024}).  
The projectile itself could have blocked part of the ejecta curtain near $\zeta = 180^\circ$, 
reducing the ejecta mass at this azimuthal angle. 

Figure \ref{fig:aves} shows that there is scatter in our ejecta mass measurements which can primarily
be attributed to variations in impact location with respect to the tray pattern. 
Scatter can also be  
attributed to density variations within the sand (though we mitigated by raking the sand prior to impact),  surface level variations of the sand prior to 
impact due to uneven leveling and disturbance caused by tray placement (less than about 1~mm), and relative variations (of order 0.3 cm) in tray locations and orientation (a few degrees) between experiments.   
%The variations in impact location are about 1 cm based on  examination of crater locations except at low impact angle where they can be larger. 
We did not carry out experiments of ejecta mass at the grazing angle of $\theta_I = 10^\circ$ because 
the crater is more elongated and smaller than at higher impact angles, 
and because it was more difficult to accurately aim the projectile to center the impact site  at low impact angles.   
%For the $\theta_I = 20^\circ$ impact angle trials, we discarded two experimental trials because the crater overlapped the tray edges.  
%Based on crater locations, we estimate that variations in impact location have a standard deviation  of about 1 cm.  

Despite the scatter in individual measurements and the low mass of ejecta measured 
in each tray (often well below 1 g per bin), Figure \ref{fig:aves} shows that azimuthal variations in ejecta mass 
are robustly measured in our experiments. 
The ratio between the ejecta mass in the downrange bin to that in the uprange bin 
is large, ranging from about 3 at $\theta_I = 80^\circ$ to about 8 at $\theta_I = 50^\circ$
(see Figure \ref{fig:amps} below).
The differences between up and downrange ejecta are so large that they exceed the errors  
introduced by centering the trays at the crater center rather than the impact point and by variations 
in impact location. 

\begin{table}[ht]\centering
\caption{Total ejecta mass measured as a function of impact angle}
\label{tab:mtot}
\begin{tabular}{lllll}
\hline
$\theta_I$ & $M_{ej}(\theta_I)$  \\ 
(deg) & (g)\\
\hline
20 & 2.27 $\pm$ 0.93 \\
%20 & 1.61 $\pm$ 1.06 \\
30 & 3.81 $\pm$ 0.32 \\
40 & 6.89 $\pm$ 0.75 \\
50 & 6.57 $\pm$ 1.12 \\
60 & 6.64 $\pm$ 0.70 \\
70 & 7.35 $\pm$ 0.60 \\
80 & 7.34 $\pm$ 1.22 \\
90 & 6.66 $\pm$ 0.81 \\
\hline
\end{tabular}
\begin{tablenotes}  
      \small \item Total ejecta mass in g as a function of impact angle. 
      Each total ejecta masses is an average from 6 experiments. 
      The uncertainty is the standard deviation computed from the
      scatter in the 6 experiments at each impact angle. 
      \end{tablenotes}
\end{table}

\subsection{Total ejecta mass as a function of impact angle}

\begin{figure}
    \centering
    \includegraphics[width=3.3truein]{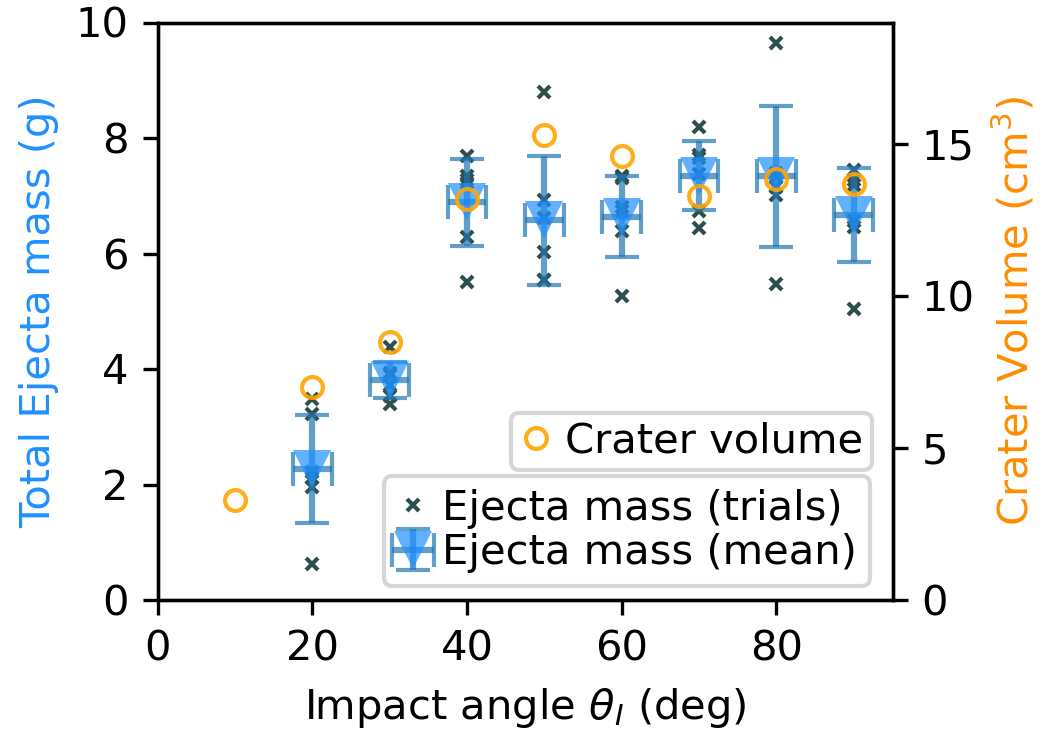} % fig6
    \caption{Total ejecta mass and ejecta volume measured at each impact angle. 
    Black x's show the ejecta mass measurements
    for each individual experiment, in g, and blue triangles show the means of these measurements, with axis
    on the left.  Crater volume measurements are shown with orange circles and with axis on the right.
    The vertical error bars are estimated from the scatter in the total ejecta mass measurements.
    The horizontal error bars represent uncertainty in impact angle. 
    }
    \label{fig:mej}
\end{figure}

Figure \ref{fig:mej} shows total ejecta mass at each impact angle, $\theta_I$.  These are computed by
summing the ejecta mass from the azimuthal bins.   
%Black x's show total ejecta mass for each experiment and the large blue triangles show the averages computed from experiments at each impact angle. 
%The vertical error bars are estimated from the scatter in the total ejecta mass measurements.
%The horizontal error bars represent uncertainty in impact angle. 
% this is correct, don't remove 
%The ejecta mass measurements are in g and have axis on the left.
%Crater volume measurements are shown with orange circles, are in cm$^3$,  and have axis on the right. 
Crater volume measurements are 
taken from Table 5 by \citet{Suo_2024} except that at $\theta_I = 80^\circ$ which we corrected after reexamining the crater topographic profile and the normal impact which we measured separately
using a photograph showing a single laser line illuminating the crater cross section. 
%With a photograph of single laser line of a normal impact crater, we measured crater volume,  rim radius and radius $R_{cr,slv}$, finding values that are  consistent with those measured at  $\theta_I = 80^\circ$. 
%We checked that  these crater volume and radius measurements are consistent with those measured  from using a photograph single laser line of a normal impact crater. 

Figure \ref{fig:mej} shows that the total mass in ejecta is approximately proportional to crater volume, 
 as expected from prior studies \citep{Housen_1983,Luther_2018}.  
The low ejecta mass at grazing impact angles (lower than $50^\circ$) is related to the reduction in crater volume that was previously noted by \citet{Suo_2024} and that was attributed to energy carried away via ricochet of the projectile.  A crater mass can be estimated 
from the product of crater volume times target substrate density; $M_{cr} = \rho V_{cr,slv}$. Crater mass is
 the mass that was originally located inside the crater and is defined with respect to the level of the surface
 prior to impact. 
 In Figure \ref{fig:mej} we have adjusted the y-axis scales so that
crater volume measurements lie on top of crater ejecta mass measurements, giving 
 ratio of ejecta to crater mass 
\begin{align}
\frac {M_{ej}(\theta_I)}{M_{cr}(\theta_I)} \approx 0.35 . 
\end{align} 

The shock physics code impact simulations by \citet{Luther_2018} predict a ratio of 
ejecta to crater mass of 0.37 for their high velocity impacts 
(see their Figure 15) with a lower value of 0.25 for their lowest impact velocity (1 km/s) simulation. 
Our value of 0.35 is consistent with this prediction, despite the difference in impact regime (our impact 
velocity is at least an order of magnitude lower).  In hypervelocity ($\sim 6 $ km/s) experiments into layered sand,  \citet{Stoffler_1975} attributed the relatively large value of crater mass 
compared to ejecta mass  to 
compactification of the target substrate and deformation associated with radial expansion and rim uplift (which 
 exceeds ejecta curtain depth \citealt{Sharpton_2014,Sturm_2016,Neiderbach_2023}). 
Our low value of the ratio of ejecta to crater mass does not necessarily imply that our experiments 
fail to capture most of the ejecta, even though our trays do not capture some low velocity material 
 ejected very near the crater rim and some downrange high velocity ejecta from low impact angle ($\lesssim 30^\circ$) impact experiments.  

\begin{figure}[ht!]
\centering
\includegraphics[width=3truein, trim = 0 25 0 0, clip]{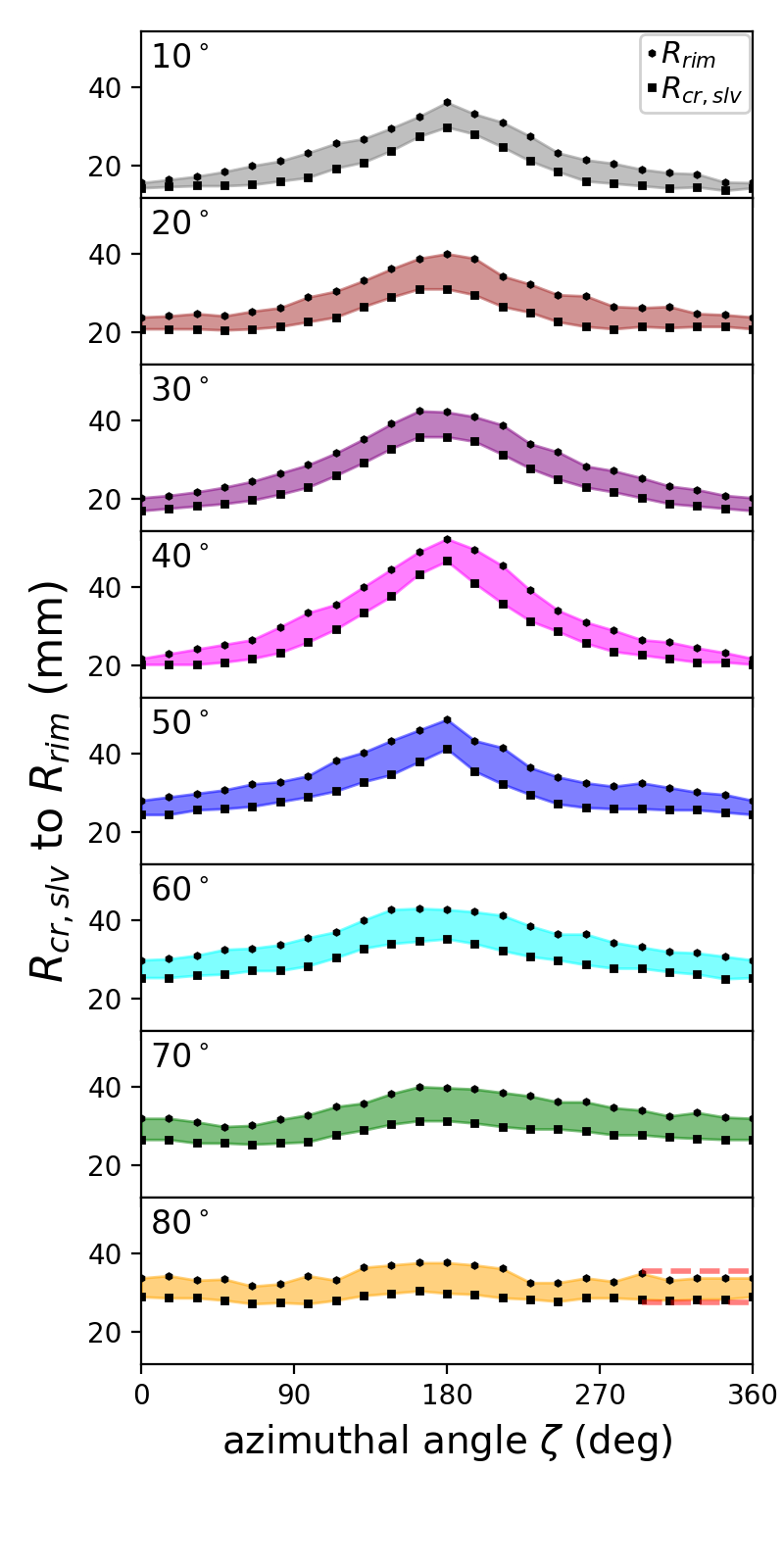} % fig7
\caption{Radius of crater rim peak, $R_{rim}$ and crater radius  
$R_{cr,slv}$ as a function of impact angle $\theta_I$ and azimuthal angle $\zeta$.  We measured these from the crater topographic profiles obtained by \citet{Suo_2024} using nearly the same experimental apparatus. 
In each panel we plot measurements at a different impact angle $\theta_I$ which is labeled in degrees
on the top left of each panel.   
The horizontal axis is the azimuthal angle $\zeta$.  
Rim peak $R_{rim}$ radii in mm from the impact point are shown
with the black circles and $R_{cr,slv}$ (the radii at which the crater depth is equal to the target substrate level prior to impact) in mm are shown with black squares.  The regions between the rim and $R_{cr,slv}$ values 
are filled with color.  The shallowest grazing impact is shown in the top panel.  }
\label{fig:R_zeta}
\end{figure}

\section{Crater radii as a function of azimuthal angle}
\label{sec:radii}

Models for the ejecta velocity and mass distribution can relate the properties of the ejecta curtain 
to the crater radius \citep{Housen_2011}.   However, for oblique impacts, the crater radius, as measured 
from the impact point,  
depends on both impact and azimuthal angles. To relate our measurements of the azimuthal ejecta 
mass distribution to models for the ejecta curtain we first measure the crater radius 
as a function of impact and azimuthal angles. 

\citet{Suo_2024} measured crater topographic profiles for oblique impacts by taking a video (at a conventional frame rate)  while slowly stepping a line-laser across the crater.  
These videos were processed to measure crater depth as a function of position at 8 different impact angles and creating eight two dimensional crater depth arrays \citep{Suo_2024}.    
Here we analyze these eight crater topographic profiles again but  
to measure crater and rim radius as a function of azimuthal and impact angles. 
We did not measure 
ejecta mass distributions from the experiments that were used to measure crater topographic profiles.
However, 
the crater profiles by \citet{Suo_2024} are of impacts with the same projectiles, fired with 
the same airsoft gun, and into the same tub containing the same sand, as the experiments presented in section \ref{sec:ejmass}. 
 
 \citet{Suo_2024} measured crater topographic profiles from 8 experiments, at impact angles
 ranging from  $\theta_I = 10^\circ$ to $80^\circ$ in increments of $10^\circ$. 
 With comparison high speed videos, \citet{Suo_2024} located the location where the 
 projectile first contacts the surface,  which we call the impact point.  The topographic profiles give depth as a function of $x,y$ on the surface plane, with origin the impact point.   
For each of the eight computed two dimensional crater depth arrays,   
we extracted the depth as a function of distance from the impact point at different azimuthal angles.  Positions along a desired angle $\zeta$ were extracted using linear interpolation from the 2d-arrays depth arrays.  The topographic profiles at each azimuthal angle were smoothed using a 1 dimensional Savinsky-Golay filter
of window length 3 mm and polynomial order 3.  
The maximum height along the smoothed profile gave the radius of the rim peak $R_{rim}(\theta_I,\zeta)$.   Within the rim we identified the radius which was at the same depth as the undisturbed surface prior to impact, $R_{cr,slv}(\theta_I,\zeta)$.  
%We refer to these radii as $R_{rim}(\zeta,\theta_I)$ and $R_{cr,slv}(\zeta,\theta_I)$.  
%We are following \citet{Housen_2011} for the definition of $R_{cr,slv}$ in the context of normal impacts. 
The resulting rim and level crossing radii are plotted in Figure \ref{fig:R_zeta} as a function of azimuthal angle 
for each of 8 impact angles $\theta_I$. 

The rim diameter and crater diameter for a normal impact listed in Table \ref{tab:airsoft_bbs} we computed
from the medians of the rim radii and the medians of the crater radii 
 at different azimuthal angles using the topographic profile from the experiment
 at an impact angle of $\theta_I= 80^\circ$. 
 Twice the rim radius differs slightly from the rim to rim crater diameter along the major
 axis of 7.1 cm previously listed in Table 5 by \citet{Suo_2024}.   
We give an updated value for crater volume in Table \ref{tab:airsoft_bbs} for a normal impact
based on the same topographic profile 
as we suspect that there was a 
minor error in the value reported in Table 5 by \citet{Suo_2024}. 
Using a photograph of a laser line illuminating a cross section of a normal impact crater, 
we measured % the crater volume, 
the crater and rim radius.  
The crater radius and rim radius are shown as red dashed lines on the bottom panel of 
Figure \ref{fig:R_zeta} and are consistent with the  measurements 
for the $\theta_I=80^\circ$ impacts.  
%The crater volume for the normal impact crater  is also consistent with that measured from the $\theta_I=80^\circ$ topographic profile. 

%In Table \ref{tab:airsoft_bbs} for a normal impact we list the  crater rim radius $R_{rim,n}$, the radius $R_{cr,slv,n}$ and the crater volume $V_{cr,slv,n}$.  These quantities are measured from the crater depth profile measured by \citet{Suo_2024}  for an impact at an  impact angle $\theta_I=80^\circ$.   

%Figure \ref{fig:RR} shows crater radii $R_{rim}$ and $R_{cr,slv}$ as images. 

%\begin{figure}[ht!]\centering
%\includegraphics[width=3 truein]{RR.png}
%\caption{Crater radii measured from crater profiles as a function of azimuthal angle and impact angle.
%The y-axes show the impact angle $\theta_I$ in degrees and the x-axis shows the azimuthal angle $\zeta$. 
%The top panel shows $R_{rim}$ in mm and the bottom panel shows $R_{cr,slv}$ in mm. }
%\label{fig:RR}
%\end{figure}

\begin{figure}[ht]\centering
\includegraphics[width=3truein, trim = 8 0 0 0, clip]{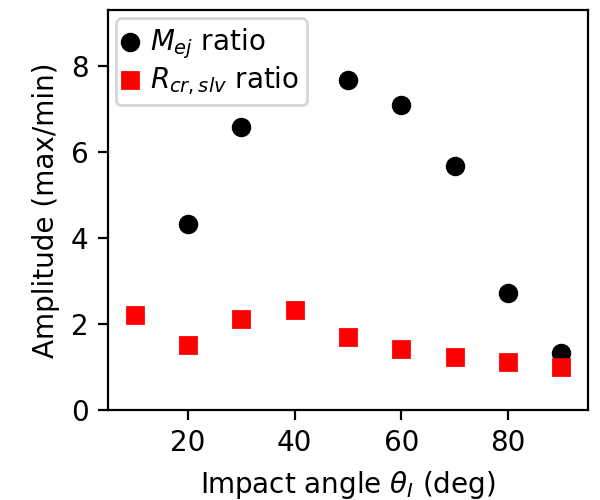} % fig8
\caption{Amplitudes of azimuthal variation in ejecta mass and crater radius as a function of 
impact angle.  With black squares we plot the ratio of the maximum  to 
minimum ejecta mass
from the different azimuthal bins using the means from 6 experimental trials at each impact angle. 
The red dots show the ratio of the maximum to minimum crater radius at each impact angle.  
The maximum values are downrange and the minimum values are uprange. 
 We do not plot the mass ratio 
 at $\theta_I = 40^\circ$ because the uprange bin had very low mean ejecta mass, giving a ratio 
above 20.  The  azimuthal variations in ejecta curtain mass are larger than azimuthal variations in crater radius. }
\label{fig:amps}
\end{figure}

The size of azimuthal variations in ejecta mass exceed those in crater radius (measured from
the impact point).  To show this, 
we examine the ratio of ejecta mass in the downrange direction to that in the uprange 
direction and we compare this ratio to the ratio of the downrange crater radius
to uprange crater radius. 
In Figure \ref{fig:amps} at each impact angle,  black dots show the ratio of the maximum 
ejecta mass to the minimum ejecta mass using the average values plotted in Figure 
\ref{fig:aves}.  The maximum values tend to be downrange and the minimum values
uprange.   At each impact angle,  with red squares,  we show the ratio of the maximum to 
minimum crater radii that are also shown in Figure \ref{fig:R_zeta}.  
Figure \ref{fig:amps} shows that the ejecta mass amplitudes of azimuthal 
variation are higher than the amplitudes of azimuthal variation in the crater radius. 

Figure \ref{fig:amps} shows that the uprange/downrange asymmetry in ejecta mass is large. 
Due to the placement of our ejecta trays (centered at crater center rather than impact point), the 
asymmetry could be even larger at low impact angles.  
If we correct by the extreme error values estimated in section \ref{sec:ejmass}),  the  ratio of maximum to minimum ejecta mass could increase by a factor as large as 1.4/0.8 = 1.7.   
%Thus, more accurate estimates of ejecta mass distributions could find that ejecta mass distributions are even more asymmetric than measured here. 
At low impact angles $\theta_I \lesssim 30^\circ$ a correction for high velocity ejecta that escaped our trays  would  increase the ratio maximum to minimum ejecta mass ratio.  A zone of avoidance that could be 
present at grazing impact angles $\theta_I \lesssim 20^\circ$ would also increase the ratio.  
% (see section \ref{sec:ejmass}). 
Misalignment between crater center and tray centers could cause asymmetry in the ejecta mass that 
is larger than that of the ejecta itself.  This might be particularly an issue for the larger craters 
at high impact angle where the crater rim is nearer the tray edges.   
Thus the ejecta mass asymmetry of the $80^\circ$ impact could have been overestimated 
but the lower impact angle asymmetries are more likely underestimated.   

Do we expect asymmetry in the ejecta mass distribution at a near normal impact at $\theta_I = 80^\circ$?
At this impact angle the ratio of downrange to uprange peak acceleration and velocity in subsurface pulses was between 1 and 2 (see Figure 13 by \citealt{Suo_2024}) and high speed video frame show 
an asymmetric ejecta curtain (see Figure 2 by \citealt{Suo_2024}).   Hence it is likely the ejecta mass 
distribution is asymmetric, but Figure \ref{fig:amps} may have overestimated the uprange to downrange ejecta mass  ratio.  

\subsection{Fitting the crater radius as a function of impact and azimuthal angles} 
\label{sec:radius}

\begin{figure*}[ht!]\centering
\includegraphics[width=2.35 truein,trim = 14 0 0 0,clip]{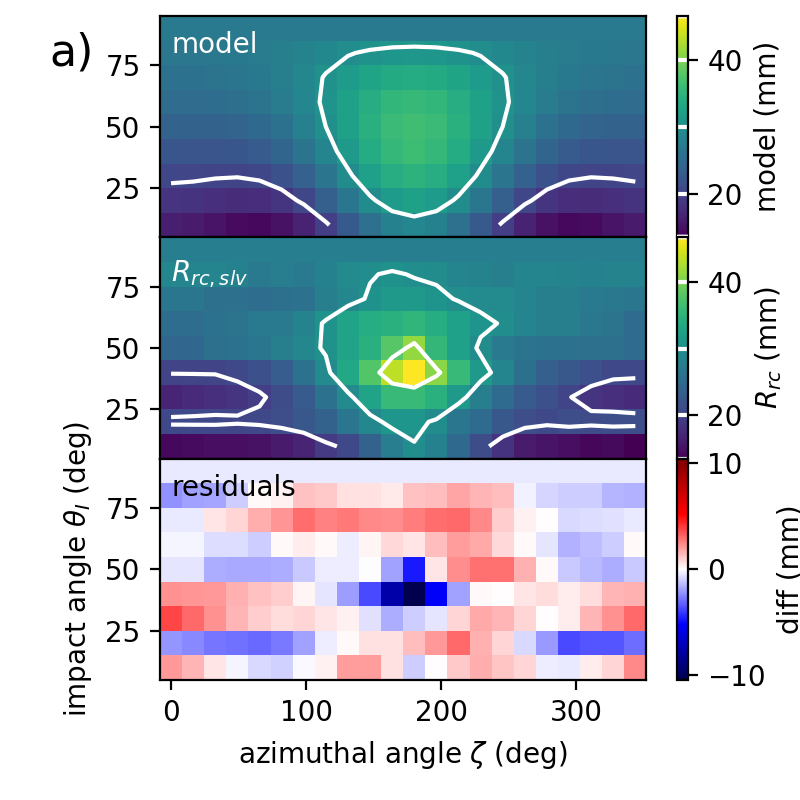} %fig9a
\includegraphics[width=2.35 truein,trim = 14 0 0 0,clip]{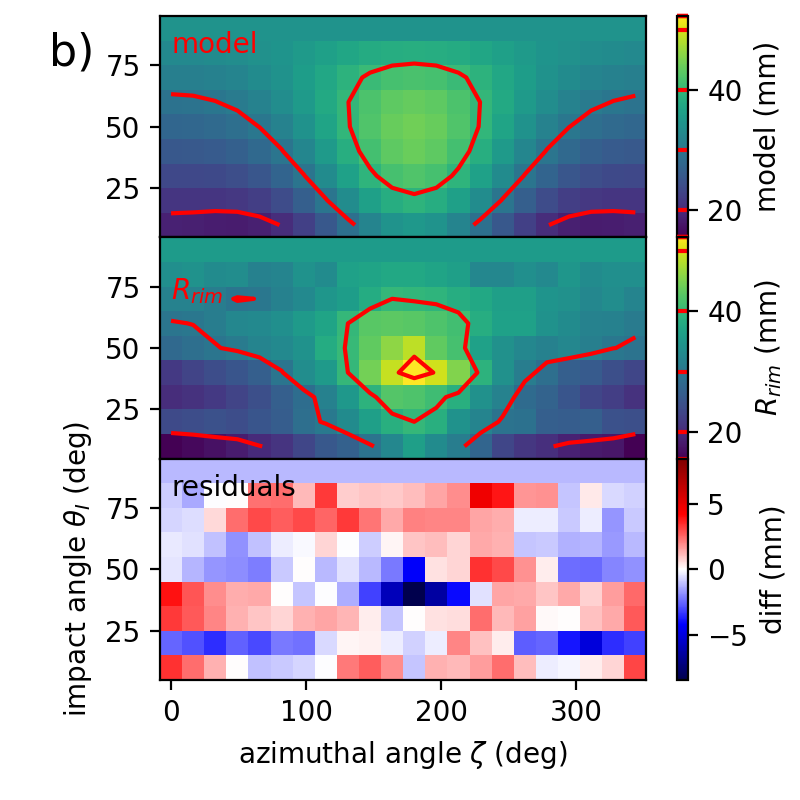} % fig9b
\includegraphics[width=2.35 truein,trim = 14 0 0 0,clip]{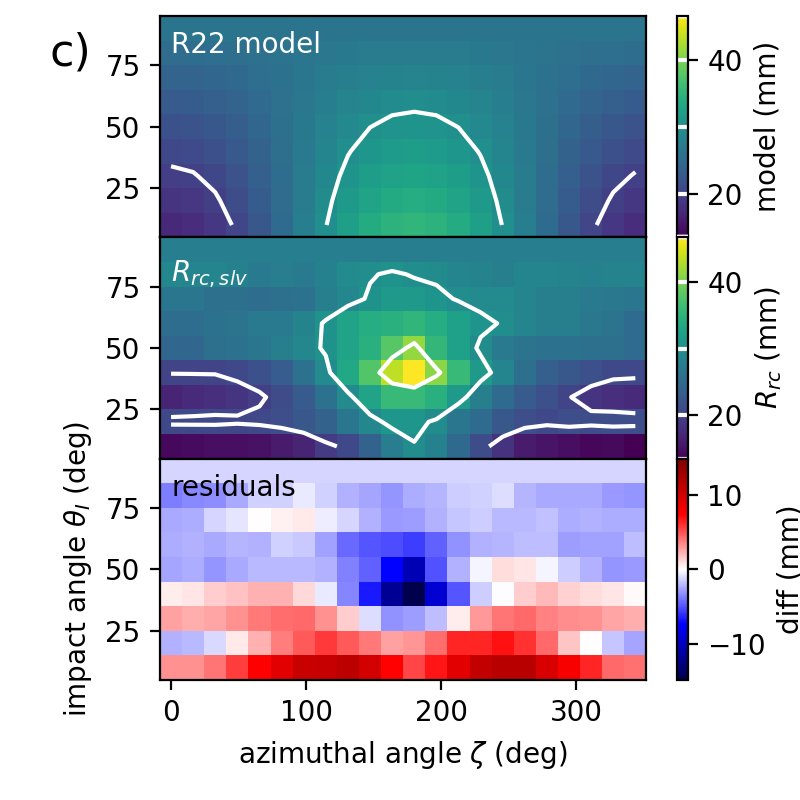} %fig9c
\caption{Results of least-squares fitting to crater radius as a function of azimuthal angle and impact angle. 
The $x$-axis is azimuthal angle $\zeta$ in degrees and the $y$-axes are impact angle $\theta_I$ in degrees. 
The color bars are in mm. 
a) The top panel shows the model of equation \ref{eqn:Rfit} 
fit to our measurements for the radius $R_{cr,slv}(\theta_I,\zeta)$.  
The middle panel shows our measurements.  These
are the same as  shown in Figure \ref{fig:R_zeta} but illustrated as a two-dimensional image.
The bottom panel shows the residuals, model subtracted by data measurements. 
b) Similar to a) except the model of equation \ref{eqn:Rfit} is fit to measurements of the rim radius $R_{rim}$.
c) Similar to a) except the model only contains 2 coefficients, $a_{00}$ and $a_{11}$ (following equation 9 by \citealt{Raducan_2022}). 
 }
\label{fig:fits}
\end{figure*}

Using least-square minimization we fit a function 
 to the radii $R_{cr,slv}(\theta_I,\zeta)$ and $R_{rim}(\theta_I,\zeta)$ 
 that we measured from the crater topographic profiles.
The function we choose extends the function for crater radius adopted 
 by \citet{Raducan_2022} that is linear in impact angle $\theta_I$ and depends on $\cos \zeta$. 
We use the function 
 \begin{align}
R(\theta_I,\zeta) =  a_{00} + \sum_{j=1}^2 \sum_{k=0}^2  a_{jk}
\left( \theta_I- \frac{\pi}{2} \right)^j   (\cos \zeta)^k 
 \label{eqn:Rfit}
\end{align}
with coefficients $a_{00}$ and $a_{jk}$ with integers $j>0$ and $k\ge 0$. 
 For the fit to crater radius $R_{cr,slv}$, the coefficient $a_{00}=R_{cr,slv,n}$ for a normal impact. 
We found that a polynomial function in $\theta_I$ required fewer coefficients for the same quality of
fit (as measured from the reduced $\chi^2$) than one that only contained trigonometric functions of $\theta_I$. 
A least squares fitting routine in python (\texttt{scipy.optimize.least\_squares} with the Trust Region Reflective algorithm) was used to find the coefficients $a_{ij}$ that minimize $\chi^2$.  Uncertainties 
 for each coefficient are 1 $\sigma$ errors that are estimated from the diagonals of the covariance matrix
 and using the variance of the residuals as an estimate for the standard deviation of each data point.  
The best fitting coefficients (via equation \ref{eqn:Rfit}) for our measurements 
of crater radius $R_{cr,slv}$ and rim radius $R_{rim}$ (shown in Figure \ref{fig:R_zeta}) are listed 
 in Table \ref{tab:fits} along with the standard deviation of the residuals.  
 in Figure \ref{fig:fits} the best fitting functions are shown in the top panels with measurements 
 in the middle panels and residuals  (data subtracted
 by model) shown in the bottom panels.  
 Figure \ref{fig:fits}a  shows our fit to crater radius $R_{cr,slv}$ measurements, and 
 Figure \ref{fig:fits}b shows our fit to $R_{rim}$ crater rim  measurements in both cases fit varying  all
 6 coefficients in equation \ref{eqn:Rfit}.  
 
\begin{figure}[ht!]\centering
\includegraphics[width=3.3truein, trim = 25 10 30 0,clip]{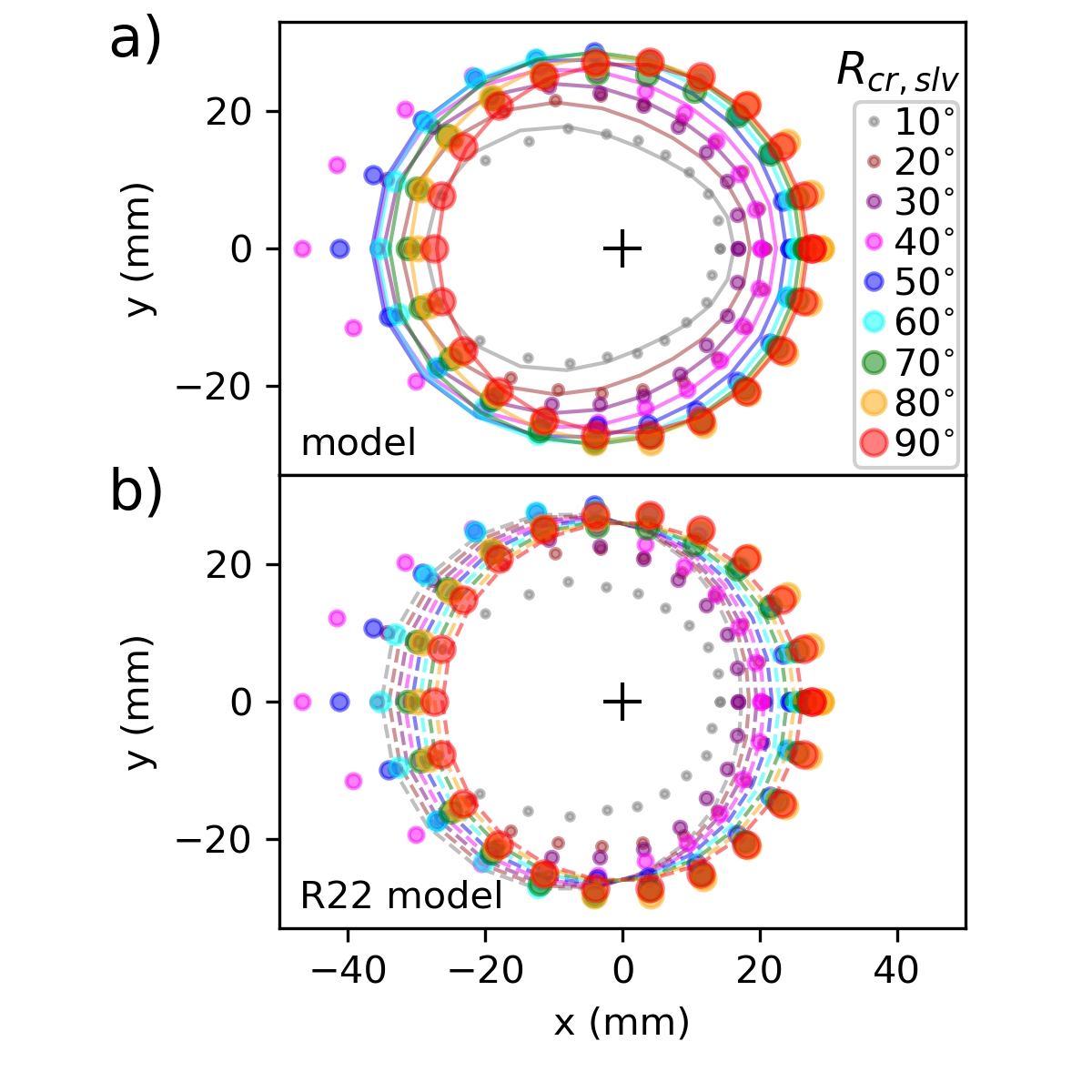} % fig10
\caption{ 
a) Circles show crater radius $R_{cr,slv}(\theta_I,\zeta)$ measured from crater topographic profiles and as seen from above
the surface.    
Each impact angle is shown with different color and size markers.
The model, described by equation \ref{eqn:Rfit}, has coefficients listed in the column labelled $R_{cr,slv}$ in Table \ref{tab:fits} and is shown with solid lines.  At each impact angle, the line color for the model matches that of the measurements that are shown with dots.  
The impact point is marked with a black cross.  The projectile originated from the $+x$ direction.  
b) Similar to a) except 
 the colored dashed lines show a two coefficient fit to the crater radius
(following \citealt{Raducan_2022}) and with coefficients listed in the column labelled $R_{cr,slv}$(R22) in Table \ref{tab:fits}.  
\label{fig:circs}
}
\end{figure}

In Figure \ref{fig:circs} the colored dots show the crater radii ($R_{cr,slv}(\theta_I,\zeta)$) we measured
as seen from above the surface. %Here the $+x$ axis is in the uprange direction. 
Dot size and color depend
upon impact angle.   In Figure \ref{fig:circs}a we plot the best fitting function (with coefficients 
for equation \ref{eqn:Rfit} and coefficients in the column labelled $R_{cr,slv}$ in Table \ref{tab:fits})
for each impact angle 
with lines, and with line color matching the dots at the same impact angle.

\begin{table}[ht!]\centering
\caption{Fits to crater radius}  
\label{tab:fits}
\begin{tabular}{lrrrcccc}
\hline
Fit to    & $R_{cr,slv}$  & $R_{rim}$   & $R_{cr,slv}$ (R22)\\
coeffs.   & (mm) & (mm)  & (mm) \\
\hline
$a_{00}$ & $ 27.03 \pm 0.32$ & $ 34.35 \pm 0.35$ & $26.23 \pm 0.29$ \\
$a_{10}$ & $  9.83 \pm 1.07$ & $  7.77 \pm 1.16$ &  \\ 
$a_{20}$ & $-11.10 \pm 0.74$ & $-10.82 \pm 0.80$ & \\ 
$a_{11}$ & $-13.64 \pm 0.91$ & $-18.24 \pm 0.98$ & $-6.36 \pm 0.50$  \\ 
$a_{21}$ & $  6.56 \pm 0.79$ & $  8.89 \pm 0.85$ & \\ 
$a_{12}$ & $  2.43 \pm 0.22$ & $  2.09 \pm 0.24$ & \\ 
\hline
Residuals\!\!\!\!\!& 1.83 & 1.97 & 4.12 \\
Figures\!\!\!     & \ref{fig:fits}a,\ref{fig:circs}a & \ref{fig:fits}b & \ref{fig:fits}c,\ref{fig:circs}b \\
\hline
\end{tabular}
 \begin{tablenotes}
      \small \item Notes:
      The $R_{cr,slv}$ column lists coefficients of the function in Equation \ref{eqn:Rfit}
       resulting from least-squares minimization fitting of our measured 
      values for crater radius $R_{cr,slv}$, shown in Figure \ref{fig:R_zeta}.  
      The $R_{rim}$ column is similar but for fits to our measured crater rim radii.  
      The column labelled $R_{cr,slv}$ (R22) shows coefficients from a fit which only
      used two coefficients $a_{00}, a_{11}$ following the function for $R_{cr,slv}$ 
      adopted by \citet{Raducan_2022}
      (their equation 9). 
The row denoted residuals list the standard deviation of the residuals in mm. 
The last row shows the figures in which measurements, model and residuals are displayed.    
 \end{tablenotes}
\end{table}

Figure \ref{fig:fits}a and Figure \ref{fig:circs}a show that our fitting function (equation \ref{eqn:Rfit}) matches
our measurements for the crater radius fairly well.  
The largest residual errors are at impact angle $\theta_I = 40^\circ$.  The elongation of the crater
along the downrange direction  at this impact angle 
can in part be attributed to the ricochet of the projectile 
that grazed the crater rim along the negative $x$ axis downrange of impact \citep{Suo_2024}.  

\citet{Raducan_2022} fit the crater radii of their impact simulations with 
a function (their equation 9)  similar to equation \ref{eqn:Rfit}, dependent upon azimuthal and impact angle,  
but only containing the two coefficients   
$a_{00} = R_{cr,slv,n}$ and  $a_{11} = - R_{cr,slv,n}\frac{1}{2} \frac{1}{100} \frac{180}{\pi} = -0.29R_{cr,slv,n}$.
%\begin{align}
%R_{cr,slv}(\theta_I,\zeta) &= R_{cr,slv,n} \left(1- \frac{ (\pi/2  - \theta_I)}{100 \pi/180}  \frac{ \cos \zeta}{2} \right) \\
%R_{cr,slv}(\theta_I,\zeta) = a_{00}  +  a_{11} \left(  \theta_I -  \frac{\pi}{2} \right)  \cos \zeta 
%\label{eqn:Radu}
%\end{align} 
%where  $\theta_I,\zeta$  are in radians and with $a_{11} = - \frac{1}{2} \frac{1}{100} \frac{180}{\pi} $. 
In Figure \ref{fig:fits}c we show a least-square model that minimizes  
a version of the function in equation \ref{eqn:Rfit} containing only two free parameters, 
$a_{00}$ and $ a_{11}$.  The resulting best fit two coefficient model and 
residuals are shown in Figure \ref{fig:fits}c.  The values 
 of the two coefficients for this fit are listed in the rightmost column in Table \ref{tab:fits}. 
The coefficients of our best fit two coefficient model has a ratio of $a_{11}/a_{00} = -0.24$ which is similar to the value of -0.29 of the model by \citet{Raducan_2022} (their equation 9). 
The residuals between the two-component model and our measurements are higher for the two coefficient model than that with six coefficients,   as would be expected since
 fewer free parameters are varied in the $\chi^2$ minimization.   Residuals between 
 our measurements and the two-component model are primarily high at low impact angle
where our crater radii are smaller than at higher impact angle.    Figure \ref{fig:circs}b,  
where we show the crater shapes predicted by the two-component model (as dashed lines) along with our crater shapes (shown with dots), also shows that our measured crater radii are smaller than predicted by the two-component model at low impact angle. 

The impact simulations by \citet{Raducan_2022} were of a 7 km/s impact.   
The function used by \citet{Raducan_2022} does not have a smaller crater at low impact angles,
which was true of our experiments.  Craters are smaller because 
 impact energy and momentum was carried away by the projectile
during ricochet at grazing impact angles in our lower velocity experiments \citep{Suo_2024}.  
The simulations by \citet{Raducan_2022} were of a high velocity (7 km/s) impact. At this velocity
the projectile could melt or fragment and at low impact angle fragments of the projectile can ricochet \citep{Gault_1978}.  
The two component function fit by \citet{Raducan_2022} to their simulated impacts for crater radius
as a function of azimuthal and impact angle gave a close fit  to the radii we measured 
 (particularly at impact angles
greater than $40^\circ$), despite the difference in impact regime. 

\section{Crater ejecta distribution models}
\label{sec:ejmodels}

\begin{figure}[ht]\centering
\includegraphics[width=2.5truein]{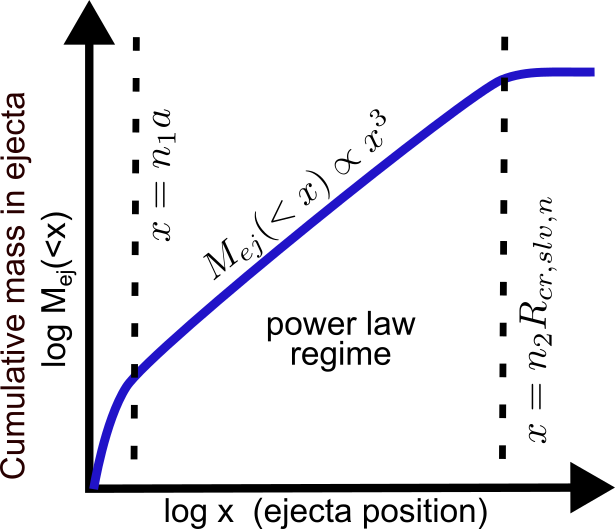} %fig11
\caption{Illustrating the point-source approximation by \citet{Housen_2011} for 
the cumulative ejecta mass distribution and the region 
where power law scaling applies.  
 \label{fig:powerlaw}}
\end{figure}

%The ejecta model for normal impacts by \citet{Housen_2011} assumes that the velocity and mass distributions for most of the ejecta are independent of projectile radius.     
The point-source approximation is the 
assumption that  ``the projectile appears as a point source when considering
any crater-related phenomena that occur very far from the impact
point''  \citep{Housen_2011}, including the ejecta velocity and mass distributions.  
%This assumption is described as the ``point-source approximation''.  
Ejecta velocity and mass 
 is a power law function of $x$, the distance between ejecta launch position and impact point. 
The power law functions are approximately truncated at an  
inner radius that is approximately equal to the projectile radius $a$ and outer radius 
that is approximately equal to the crater radius  $R_{cr,slv}$, 
%(shown in Figure \ref{fig:crater_dims}a and that we measured in section \ref{sec:radii}).  
as illustrated in Figure \ref{fig:powerlaw}.
The ratio of crater to projectile radius in our experiments is 
 $R_{cr,slv,n}/a = 9.6$. 

The point-source approximation has been experimentally verified for impacts between  
about 150  m/s and a few km/s, and holds 
as close as 1 projectile radius from the impact point \citep{Holsapple_1987,Holsapple_1993,Housen_2011}.
The point source approximation is expected to hold for impact speeds above the target substrate sound speed 
\citep{Housen_2011}.  We measured a pulse travel speed of about 55 m/s in our sand target medium  
 \citep{Quillen_2022,Suo_2024}, so our experiments satisfy this limit. 
 
The ejecta scaling model by \citet{Housen_2011} has total mass ejected at all positions up to $x$ 
from the impact point 
\begin{align}
    M_{ej,n}(<x) &= k_n \rho ( x^3  - n_1 a^3)  \nonumber \\
    & \quad {\rm for} \quad n_1 a\! < \!x\!<\! n_2 R_{cr,slv,n}.
    \label{eqn:M}
\end{align}
This power law form for the ejecta mass distribution of a normal impact is illustrated in Figure \ref{fig:powerlaw}.  
The parameter $ k_n$ is dimensionless but can depend upon the material properties \citep{Housen_2018}.  \citet{Housen_2011} found that parameters $n_1 \approx 1.2$ and $n_2 \approx 1$. 

Following \citet{Raducan_2022}, we modify the coefficients used in the crater ejecta scaling 
laws by \citet{Housen_2011} to become functions of impact and azimuthal angles $\theta_I,\zeta$.  
We modify equation \ref{eqn:M} to consider the ejecta mass in an azimuthal bin of size $d \zeta$ in radians 
\begin{align}
dM_{ej}(<\!x,\theta_I, \zeta) d\zeta &=\! \frac{\rho}{2 \pi } k(\theta_I,\zeta) ( x^3  - n_1 a^3) d \zeta \nonumber  \\
& \ \ \  {\rm for} \ \   n_1 a\! < \! x\!<\!n_2 R_{cr,slv}(\theta_I, \zeta).
\label{eqn:dMej}
\end{align}
The factor of $2 \pi$ arises because the total ejecta mass is found by integrating
over the azimuthal angle $\zeta$.   The function $k(\theta_I,\zeta)$  
has limit 
\begin{align}
\lim_{\theta_I \to \pi/2} k(\theta_I,\zeta)   = k_n
\end{align}
where $k_n$ is the coefficient for a normal impact in equation \ref{eqn:M}. 
The crater radius is dependent upon azimuthal and impact angles and 
\begin{align}
\lim_{\theta_I \to \pi/2} R_{cr,slv}(\theta_I,\zeta)  = R_{cr,slv,n}
\end{align}
where $R_{cr,slv,n}$ is the crater radius for a normal impact in equation \ref{eqn:Rcr_grav}. 
We assume that coefficients $n_1$ and $n_2 $ are independent of 
impact and azimuthal angle. 

The total ejecta mass (integrated out to the crater radius) in an azimuthal bin of width $d\zeta$
in radians is 
\begin{align}
dM_{ej} (\theta_I, \zeta) \approx  \frac{\rho}{2 \pi } k(\theta_I,\zeta) R_{cr,slv}^3(\theta_I,\zeta)  d \zeta
\label{eqn:dM}
\end{align}
where we have used $n_2 \approx 1$ found by \citet{Housen_2011} in equation \ref{eqn:dMej} for the
maximum value of ejecta radius and assumed that crater radius is much larger than the projectile radius; 
$R_{cr,slv,n} \gg a$.  %Equality in equation \ref{eqn:dM} is reached   when crater radius is much larger than the projectile radius; $R_{cr,slv,n} \gg a_{pj}$. 

\begin{figure}[ht]\centering
\includegraphics[width=3.3truein, trim = 10 0 0 0,clip]{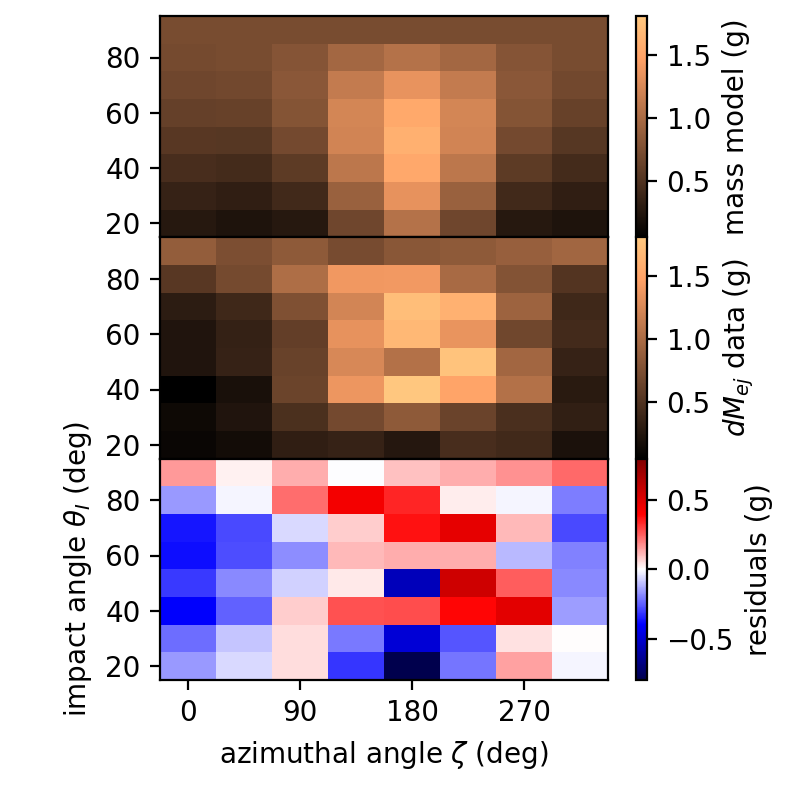} % fig12
\caption{Ejecta mass as a function of impact and azimuthal angle. 
The top panel shows the best fit model from equation \ref{eqn:dM} with function equation \ref{eqn:kfun} 
for the function $k(\theta_I, \zeta)$ and parameters in equation \ref{eqn:knk1}.  The model uses crater radius model for $R_{cr,slv}(\theta_I,\zeta)$
discussed in section \ref{sec:radius} and shown in Figure \ref{fig:fits}. 
The middle panel shows the measurements of ejecta mass in $45^\circ$ wide azimuthal bins.
Each pixel consists of an average of 6 experimental data points. 
The bottom panel shows residuals (data subtracted by model). }
\label{fig:Mmodel}
\end{figure}

For the function $k(\theta_I,\zeta)$, we adopt a form for the function;  
\begin{align}
k(\theta_I,\zeta) = k_n  + k_1 \cos \theta_I \cos \zeta .
\label{eqn:kfun}
\end{align}
Using equation \ref{eqn:dM} and parameters based on our model for the crater radius 
$R_{cr,slv}(\theta_I,\zeta)$ 
(described in section \ref{sec:radius}) 
we fit $k_n$ and $k_1$ to the mean ejecta mass  
we measured as a function of
impact and azimuthal angle.  The averages at each impact angle
consistent of mean values from measurements 
of 6 experiments per impact angle and at each of 8 possible azimuthal bins. 
Our least-squares minimizing fitting procedure is the same as 
described in section \ref{sec:radius}. The model that minimizes $\chi^2$ is shown in 
Figure \ref{fig:Mmodel}.  
Best fitting parameters for the function in equation \ref{eqn:kfun} are 
\begin{align}
k_n &= 0.19 \pm 0.01 ~{\rm g} \nonumber \\
k_1 &= 0.03 \pm 0.02  ~{\rm g}   \label{eqn:knk1}
\end{align}
with a standard deviation of 0.27 g in the residuals (model subtracted by measurements)
that are shown in the bottom panel in Figure \ref{fig:Mmodel}.   

Our best fitting value for $k_n$  is at the lower end of the range 
0.2 to 0.5 for normal impacts into target substrate materials inferred by \citet{Housen_2011} 
via comparison of their ejecta scaling model to experimental measurements.  
Our trays do not catch all the lowest velocity ejecta, which may in part account
for our low $k_n$ value. 
%Discussion:  $k_n$ is  low.  
%  \citet{Yamamoto_2005} got 0.47 for impacts into a substrate consisting of glass spheres. 
% Yamamoto, S., Okabe, N., Sugita, S., Matsui, T., 2005b. Measurements of ejecta velocity distribution by a high-speed video camera. Lunar Planet. Sci. XXXVI. Abstract 1600.  Was very low velocity!

Our best fitting value for $k_1$ is near zero,  which implies that the function $k(\theta_I,\zeta)$ is only weakly dependent upon impact and azimuthal angles. Variations in the total ejecta mass (integrated in ejection launch position $x$) as a function of azimuthal angle in equation \ref{eqn:dM} are primarily due to the sensitivity of the crater radius to impact and azimuthal angles.  
 
The function used by  \citet{Raducan_2022} to describe azimuthal variations in the ejecta mass, 
 $k(\theta_I,\zeta) = \frac{k_n}{2\pi} \exp{ \left[-0.02\cos(\zeta)\cos(\theta_I)\right]}$
(their equation 8, which we have modified to be per bin in $\zeta$).  This function is 
 weakly dependent upon impact and azimuthal angles. 
Our measurements of the ejecta mass azimuthal mass distribution 
confirm this weak dependence.   We suspect that Figure 12c by \citet{Raducan_2022} is probably not
showing their function $k(\theta_I, \zeta)$ (given in their equation 8) 
as the plotted function is more sensitive to azimuthal and impact angles than predicted by
their function.   Figures 12a, b and d by \citet{Raducan_2022} may  
have flipped or mislabelled x-axes as the plotted functions appear inconsistent with their equations 6, 7 and 9.  
Our experimental measurements are in agreement with two of the functions proposed by \citet{Raducan_2022};  
their equation 8 (for $k(\theta_I,\zeta)$) and their equation 12 (for $R_{cr,slv}(\theta_I,\zeta)$), despite some confusion caused by their Figure 12. % showing these functions.  

%\url{https://ui.adsabs.harvard.edu/abs/2019Icar..325...67B/abstract}
%Impacts into coarse-grained spheres at moderate impact velocities: Implications for cratering on asteroids and planets, Barnouin+19, not cited, velocities are higher than ours, focus is coarseness of medium

%\url{https://ui.adsabs.harvard.edu/abs/2015Icar..262...79T/abstract}
%Ejecta velocity distribution of impact craters formed on quartz sand: Effect of projectile density on crater scaling law , Tsujido+15, also not cited.  

\section{Discussion and Summary}

In this study we have measured the azimuthal ejecta mass distribution for $\sim 104$ m/s velocity 
oblique impacts into sand.   Our measurements extend our characterization of  $\sim 100$ m/s oblique impacts in a granular system, building upon the experimental study by \citet{Suo_2024}. 
Using paper trays to catch ejecta, 
we have measured the azimuthal ejecta mass distribution as a function of impact angle.
Even though the craters are nearly round, 
the ejecta curtains are strongly asymmetrical, with more ejecta launched in the downrange than
uprange direction, confirming the study by \citet{Hessen_2007} who reported ejecta asymmetry 
from 244-260 m/s impacts into sand at impact angles up to $\theta_I = 60^\circ$.   
The ratio of ejecta mass in the downrange bin to the uprange one is large and ranges from a few
at an impact angle of $\theta_I = 70^\circ$ to at least 8 at an impact angle of $\theta_I = 40^\circ$. 
%The crater center is not located at the same position as the impact point. 
The large ejecta mass in the downrange direction is related to  the strong, high   
amplitude subsurface pulses on the downrange side measured with embedded accelerometers by \citet{Suo_2024}.   

Crater rims from oblique impacts are not centered about the impact point. 
Using crater topographic profiles measured by \citet{Suo_2024} we have measured 
the crater radius $R_{cr,slv}(\theta_I,\zeta)$ (a distance from
the impact point illustrated in Figure \ref{fig:crater_dims}a),  
as a function of impact and azimuthal angles.   
We find that the azimuthal ejecta mass distribution is approximately proportional to the cube of crater radius 
$R_{cr,slv}(\theta_I,\zeta)$.   This implies that the function $k(\theta_I,\zeta)$ 
(defined in equation \ref{eqn:dMej}),  is 
approximately constant. This function relates crater radius to ejecta mass and is a generalization 
of ejecta scaling laws derived with the point source approximation for normal impacts \citep{Housen_2011}. 
Because pulses travel slowly in our sand target substrate (at about 55 m/s), the impact velocity for our experiments  ($\sim 100$ m/s) exceeds the pulse travel speed, and so satisfies a possible requirement for the 
validity of the point source approximation \citep{Housen_2011}. 
The insensitivity of the 
function $k(\theta_I,\zeta)$ describing the ejecta mass distribution to impact and azimuthal angles facilitates 
computing quantities associated with the cumulative affects of impacts on small bodies, 
including angular momentum drain \citep{Dobrovolskis_1984,Luniewski_2024} 
and erosion \citep{Quillen_2024}.   

The shape of subsurface pulses measured by \citet{Suo_2024} from accelerometers at particular subsurface locations  
were remarkably similar for oblique impacts at different impact angles (see their Figure 17). 
\citet{Suo_2024} suggested that it might be possible to describe 
the subsurface pulses and the resulting ejecta velocity field by rescaling the amplitude of the velocity field.  
Our finding, that the ejecta mass azimuthal distribution 
depends primarily on crater radius, supports this proposal.  There may be a way to characterize 
 ejecta distribution and impact generated subsurface motions via an angular and time dependent function 
 that also gives the angular dependent crater radius function $R_{cr,slv}(\zeta,\theta)$.  
 
\citet{Raducan_2022} modified ejecta scaling laws to describe simulated 6.5 km/s oblique impacts 
 with four angular dependent functions.  
We have experimentally characterized two of these functions:  crater radius $R_{cr,slv}(\theta_I,\zeta)$ and the function $k(\theta_I,\zeta)$, defined in equation \ref{eqn:dMej}, 
though in a lower impact velocity regime ($\sim 100$ m/s) and for a granular target substrate, sand.  
Our angular dependent functions for crater radius and 
$k(\theta_I,\zeta)$ describing the ejecta mass distribution are remarkably similar to those presented by
\citet{Raducan_2022}.   Nevertheless care should be taken when applying the results of our study to interpret impacts at higher or lower impact velocities than those of our experiments. 

We found that the azimuthal asymmetry in ejecta mass was so large that we could detect it  
with our unsophisticated impact experiments.   The biggest source of error in our experiments is likely 
is due to variations in projectile aim.  Impact experiments with more accurate targeting could confirm  
and improve upon our measurements. 
As our results may not apply to higher or lower velocity impact regime, 
 the ejecta mass distribution from 
oblique impacts at different impact velocities should be carried out prior to using 
the results of our study to interpret phenomena associated with different velocity regimes.   

In future work, we also hope to experimentally characterize 
two additional functions of impact and azimuthal angles, $\theta_I$ and $\zeta$, for the coefficient $C_1$ and exponent $\mu$ used by \citet{Housen_2011} to describe the ejecta velocity as a function of ejecta launch position.   Better characterization of these functions could probe whether butterfly pattern ejecta blankets 
that are seen in high velocity (few km/s) impacts \citep{Gault_1978,Luo_2022} 
are also caused by lower velocity impacts. 
We would also like to characterize 
functions describing angular variations in ejecta angle as ejecta angle is also sensitive 
to azimuthal and impact angles and ejecta launch position \citep{Anderson_2004,Raducan_2022,Suo_2024}.

%\citep{Allibert_2023} discuss modifying scaling laws to take into account supersonic to subsonic transition.  
% Threaten to reduce errors?

{\bf Acknowledgements:} We thank Bingcheng Suo for helping us set up the experiments 
and Jim Saporito for helping us with the milling machine in the shop. 
This work has been supported by NASA grant 80NSSC21K0143. 

{\bf Data availability: } 
The data that support the findings of this study and code used to generate all figures in this manuscript are openly available at \url{https://github.com/aquillen/azimuthal_mass}.  

\bibliographystyle{elsarticle-harv}
\bibliography{refs}

\end{document}